\documentclass[a4paper,10pt,twocolumn]{article}
\usepackage{graphicx}
\usepackage{amsmath}
\usepackage{amssymb}
\usepackage{hyperref}
\usepackage{tikzscale}
\usepackage{pgfplots}
\usetikzlibrary{calc}
\usepackage[tiny]{titlesec}

\titleformat*{\subsection}{\normalsize\itshape}
\titleformat*{\subsubsection}{\normalsize\itshape}

\renewcommand{\d}{\partial}
\renewcommand{\vec}[1]{\boldsymbol{#1}}
\newcommand{\mat}[1]{\boldsymbol{#1}}
\newcommand{\dims}[2]{\genfrac{}{}{0pt}{}{#1}{#2}}
\newcommand{\del}{\nabla}
\newcommand{\Lim}[1]{\raisebox{0.5ex}{\scalebox{0.8}{$\displaystyle \lim_{#1}\;$}}}
\newcommand{\argmin}[1]{\raisebox{0.5ex}{\scalebox{0.8}{$\displaystyle \arg\min_{#1}\;$}}}
\newcommand{\argmax}[1]{\raisebox{0.5ex}{\scalebox{0.8}{$\displaystyle \arg\max_{#1}\;$}}}
\newcommand{\sign}{\textrm{sgn}}

\begin{document}

\twocolumn[{%
\begin{@twocolumnfalse}

\title{The Lie Detector}

\author{A. Young$^1$$\qquad$A. G. W. Lawrie$^2$ \\ \\
$^1$\small{\emph{Gates-Thomas Laboratory, California Institute of Technology, Pasadena, CA 91125, USA}} \\ 
$^2$\small{\emph{Hele-Shaw Laboratory, University of Bristol, Queen's Building, University Walk, Bristol, BS8 1TR, UK}}}
\date{}

\maketitle

\begin{abstract}
How many free variables do we really need to build a credible model of a physical system? Currently there is no systematic approach; we appeal to some physical principles, tune free variables by comparing with canonical cases, and hope our real-world applications interpolate between them. In this work we combine two pioneering and entirely disparate pieces of mathematics: the century-old techniques of Sophus Lie for solving differential equtions and recent work initiated by Field's medallist Terence Tao on converting NP-complete combinatorical problems into neighbouring convex optimisations. We present a novel and fully systematic procedure for designing models of physical systems with necessary and just-sufficient complexity, in marked contrast with the approach to function approximation taken by neural networks and other current approaches to machine learning. Our methodology replaces the \emph{ad-hoc} development of models to recover structure and understanding from observational, experimental or simulated data. At its core, our method seeks to find invariant properties of differential equations known as \emph{Lie symmetries}, and for this reason we have called our algorithm the \emph{Lie Detector}.  
\end{abstract}

\end{@twocolumnfalse}
}]%

\section{Introduction and motivation}

Real-world physical systems are often locally unpredictable and contain infeasibly many degrees of freedom, but a remarkable property of many complex systems is the emergence at larger scales of simpler structural phenomena that may be detected in statistical measurements. Structure, in this sense, is a persistent property with lower dimensionality than that of the underlying state of the system. It follows that there will exist a corresponding manifold of modest dimension near which all states of the system lie, and evolution of the state produces trajectories on and around this manifold.

We seek a criterion for determining how many dimensions are required describe such a manifold, and a procedure for determining from observational data what its shape should be. One might call this process \emph{model discovery}, since the dimensionality of the statistical manifold is the number of variables required, and its shape in space determines the functional form of the model.

Engineering modelling of complex systems is ubiquitous, but guarantees of model quality are not, and currently models are created \emph{ad-hoc} by invoking plausible physical arguments. Our methodology replaces \emph{ad-hoc} development with a fully systematic procedure for finding structure in observational, experimental or simulated data. 

The following section outlines our strategy for systematic model discovery. Since we anticipate that the utility of our scheme will extend far beyond the field of applied mathematics, especially into areas currently addressed using techniques of `machine learning', we proceed by offering a first-principles description of the main components of the problem: firstly recasting Lie's original framework \cite{lie1874} for solving differential equations in a numerical context, highlighting the embedded problem of determining the dimensionality of the low-dimensional statistical system, and then presenting a systematic solution to this embedded problem. We then discuss implementation details, present results from a proof-of-concept test case and provide a link to a demonstration code.

\section{Outline of strategy}

We make the assumption that any data-set we seek to understand, whether observational, experimental or simulated, has arisen from a problem for which there is some underlying coherence. Random, uncorrelated and incoherent data is devoid of structure and thus has no pattern we could hope to discover. However, beneath all behaviour arising from physics is some form of coherence, and this is often best described by conservation laws and associated differential equations. Conservation statements may also be described as \emph{invariant properties} of a system, or as \emph{symmetries}. The method we present here is a strategy for discovering such invariant (or near-invariant) properties, and expressing the corresponding differential equation. 

\subsection{Similarity solutions}

Solutions to differential equations that by a suitable rescaling of coordinates preserve their own structure are categorised as \emph{self-similar}. A standard \emph{ad-hoc} solution approach uses some prior expectation of the rescaling, leaving only a small number of free parameters to determine a general solution. The rule for rescaling is an invariant property of the equation - the essence of its structure. Essentially all techniques for solving differential equations reduce to the pursuit of an invariance.   

\subsection{Lie Groups}

The standard solution to an integration problem $\int f'(x) dx$ produces a family of solutions $f(x)+c$, each differing by an integration constant, $c$. Sophus Lie had the realisation, in 1874, that the flexibility of the solution afforded by adding `$+c$' was in fact a form of transformation that has a profound connection to group theory, and published his work in a series of monographs \cite{lie1874} between 1888 and 1893. A comprehensive modern reference, \cite{olver}, covers developments in the intervening century. The group of transformations takes an invariant structure $f(x)$ and slides it around, with an identity element at $c=0$, and all the required properties of a group even although it contains uncountably many elements. 

With suitable coordinate transformations, much more general classes of differential equation could be reformulated as an invariant structure and a rule for transforming it through coordinate space, and from this Lie established a systematic process for hand-calculating their solutions. The coordinate transformation lies at the heart of Lie's method, and rather than impose an \emph{ad-hoc} prior expectation on its form, Lie proposes using a polynomial series expansion to form a linear system whose solution determines the ideal coordinate transformation. 

\begin{figure}
\centering
\begin{tikzpicture}

\begin{axis}[clip=false, xmin=-2.0, xmax=2, ymin=-2, ymax=2.4, xlabel=$x$, ylabel=$f(x)$, width=\linewidth, at={(0.52\textwidth,0)}, axis x line=middle, axis y line=middle, height=0.5\linewidth, xticklabel style={below, yshift=2pt}, anchor={south east}]
\addplot[domain=-2:2, black, thick] {0+(x^3/3)+(x^2/2)-x};

\addplot[domain=-2:2, blue] {0.2+(x^3/3)+(x^2/2)-x};
\addplot[domain=-2:2, blue] {-0.2+(x^3/3)+(x^2/2)-x};
\addplot[domain=-2:2, blue] {0.4+(x^3/3)+(x^2/2)-x};
\addplot[domain=-2:2, blue] {-0.4+(x^3/3)+(x^2/2)-x};
\addplot[domain=-2:2, blue] {0.6+(x^3/3)+(x^2/2)-x};
\addplot[domain=-2:2, blue] {-0.6+(x^3/3)+(x^2/2)-x};
\addplot[domain=-2:2, blue] {0.8+(x^3/3)+(x^2/2)-x};
\addplot[domain=-2:2, blue] {-0.8+(x^3/3)+(x^2/2)-x};

\draw[-latex, thick, ->] (axis cs:-2.05,1.1) -- (axis cs:-2.05,0.2) node[left,midway] {$c<0$};
\draw[-latex, thick, ->] (axis cs:-2.05,1.3) -- (axis cs:-2.05,2.3) node[left,midway] {$c>0$};
\draw[-latex, thick, ->] (axis cs:2.05,2.6) -- (axis cs:2.05,1.7);
\draw[-latex, thick, ->] (axis cs:2.05,2.8) -- (axis cs:2.05,3.8);

\end{axis}
\end{tikzpicture}

\caption{Solutions to $\int f'(x) dx$ are of the form $f(x)+c$. These solutions form a family, known as a transformation group, with an associative group operation $+$, an identity ($c=0$), inverse ($-c$), and since all transformations result in a member of the group, closure is assured. Shown are some members of the family for $f'(x)=x^2+x-1$, with the identity in black.
}
\label{fig:transformationgroup}
\end{figure}
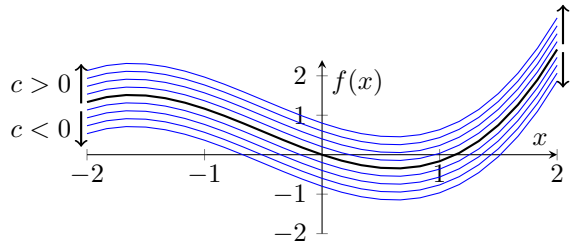

In our numerical implementation we seek to discover functions $f$ given by,
\begin{equation}\label{eqn:intronontrivial}
\frac{dx}{dt}=f(x,t).
\end{equation}
By considering such equations as surfaces in the three-dimensional space of $\{t,x,\dot{x}\}$, we may utilise the orthogonality of surface normals and vectors tangent to the plane. Expressing tangent vectors as a polynomial series expansion, we form a linear system with matrix, $\mat{B}$, where each of its columns represents one of the polynomial basis functions, and each row is a coordinate in $\{t,x,\dot{x}\}$ space at which we evaluate the polynomial. We obtain a set of coefficients $\vec{\eta}$ for which $\mat{B}\vec{\eta}=\vec{0}$, from the scalar product of normals with tangents. 
 
The existence of a unique `$+c$' direction in which solutions can be transformed whilst preserving an invariant structure relies on the matrix being rank-deficient, but only slightly. A choice of basis functions that produces a null space (the set of $\vec{\eta}$ for which $\mat{B}\vec{\eta}=\vec{0}$) of one dimension guarantees that there is only one degree of freedom in which solutions may then be transformed, and this corresponds to the $+c$ direction. The series of basis functions is typically expanded to successively higher orders until a null-space emerges. 

Conventionally $f$ would be a known function and we would seek an invarant direction to find explicit solutions to (\ref{eqn:intronontrivial}). For given $f$, one may determine analytically when the polynomial basis is expanded sufficiently to produces a null space, and the coordinate transformation aligned with the $+c$ direction condenses the problem into fewer variables, which may be simpler to solve directly. 

\begin{figure}
\centering
\begin{tikzpicture}

\node[anchor=north west] (xlabel) at (4.0,0) {$\vec{\eta}_1$};
\node[anchor=south west] (ylabel) at (0,3.5) {$\vec{\eta}_2$};
\draw[->] (-2.5,0.0) -- (xlabel.north west);
\draw[->] (0.0,-2.5) -- (ylabel.south west);

\def\pw{0.05};
\def\mw{-0.05};
\def\psz{2.0};
\def\msz{-2.0};
\definecolor{BlackPoint}{rgb}{0.0,0.0,0.0}
\draw[BlackPoint] (\msz,0.0) -- (\mw,\pw);
\draw[BlackPoint] (\msz,0.0) -- (\mw,\mw);
\draw[BlackPoint] (0.0,\msz) -- (\pw,\mw);
\draw[BlackPoint] (0.0,\msz) -- (\mw,\mw);
\draw[BlackPoint] (0.0,\psz) -- (\pw,\pw);
\draw[BlackPoint] (0.0,\psz) -- (\mw,\pw);
\draw[BlackPoint] (\psz,0.0) -- (\pw,\pw);
\draw[BlackPoint] (\psz,0.0) -- (\pw,\mw);

\definecolor{BluePoint}{rgb}{0.0,1.0,0.0}
\draw[thick,BluePoint] (0.0,\psz) -- (\psz,0.0);
\draw[thick,BluePoint] (0.0,\msz) -- (\psz,0.0);
\draw[thick,BluePoint] (0.0,\msz) -- (\msz,0.0);
\draw[thick,BluePoint] (0.0,\psz) -- (\msz,0.0);
\node[anchor=west] (L1) at (0.05,2.1) {$L_0, L_1$};
\fill[BluePoint] (0.1,\psz) arc (0:180:0.1);
\fill[BlackPoint] (-0.1,\psz) arc (180:360:0.1);

\definecolor{GreenPoint}{rgb}{0.0,0.0,1.0}
\draw[GreenPoint] (1.84,0.0) arc (0:360:1.84);
\draw[GreenPoint] (0.0,0.0) -- (0.63,1.68);
\node[anchor=west] (L2) at (0.73,1.72) {$L_2$};
\fill[GreenPoint] (0.75,1.76) arc (0:360:0.1);

\definecolor{RedPoint}{rgb}{1.0,0.0,0.0}
\def\pli{1.44};
\def\mli{-1.44};
\draw[RedPoint] (\mli,\mli) -- (\mli,\pli);
\draw[RedPoint] (\mli,\pli) -- (\pli,\pli);
\draw[RedPoint] (\pli,\pli) -- (\pli,\mli);
\draw[RedPoint] (\pli,\mli) -- (\mli,\mli);
\draw[RedPoint] (0,0) -- (\pli,\pli);
\node[anchor=west] (Li) at (1.52,1.44) {$L_{\infty}$};
\fill[RedPoint] (1.555,1.455) arc (0:360:0.1);

\node[anchor=south west,rotate=-21] (nullspace) at (-2.5,2.8) {1D null space};
\draw[dashed] (-2.5,2.9375) -- (4.0,0.5);  

\end{tikzpicture}

\caption{For a simple $2\times 2$ system $\mat{B}\vec{\eta}=\vec{0}$ that is rank-deficient by one, there is no unique solution $\vec{\eta}$, but all lie somewhere on a line known as the null space. To identify one solution as optimal in some sense, we typically seek to minimise an $L_p$ norm of the solution vector. The figure shows the $L_p$ `ball' for $p\to0$, $p=1$, $p=2$, $p\to\infty$. The $L_0$ and $L_1$ optimal solutions are identical and $p>1$ optima lie off the coordinate axes and thus involve more non-zero components. This approach can be shown (\cite{candesrombergjustintao2006a,candesrombergjustintao2006b}) to generalise to $n$-dimensions.}
\label{fig:lpball}
\end{figure}
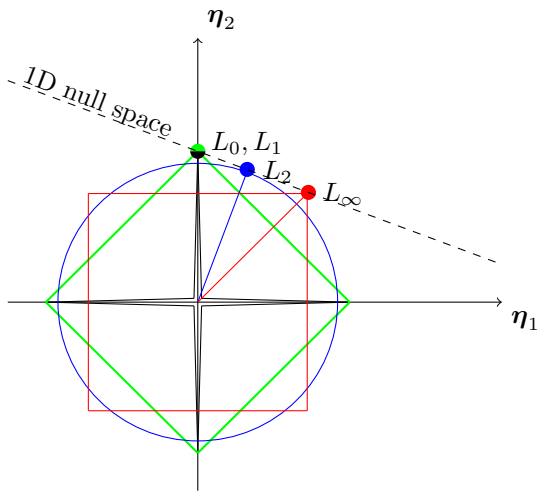

\subsection{Determination of matrix rank}

The key insight behind this work is that we recognise that Lie's method is fully reversible in principle, and can be performed backwards from samples of particular solutions (such as observed, experimental or simulation data) to recover an equation that succinctly describes the structure of that solution. The practical problem of doing so lies in guaranteeing the dimensionality of the null space. Contamination by noise will result in matrices that appear to be full rank to numerical precision, even when the underlying structure may not be.

\subsection{Generalised norms} 

Work in an entirely unrelated field, initiated in 2006 by Tao in a series of papers, \cite{taocandes2006,candesrombergjustintao2006a,candesrombergjustintao2006b}, provides a foundation for implementing this guarantee. In general terms, decision-based problems, whose solution is determined by existence \emph{vs.} absence rather than the value of a variable, cannot be solved without a brute-force scan of the permutations. Such problems are classified as NP-complete. Tao offered a new way of viewing certain classes of NP-complete problems as minimisations of the norm of a vector that indicates state in the problem. The norm of some vector $\vec{x}$ is defined as, 
\begin{equation}
\left\| \vec{x} \right\|_p = L_p(\vec{x})=\left(\sum\limits_i |\vec{x}_i|^p\right)^{\frac{1}{p}},
\end{equation} 
In the limit $\Lim{p\to 0}$, the $L_0$ norm represents the number of non-zero values in the vector. A solution to a problem that minimises the $L_0$ norm can be thought of as the least-complex solution the problem (the \emph{Occam's Razor} solution) rather than the one that minimises the vector length (or `energy', or `Gaussian variance' of the solution, depending on the context), which is obtained from the $L_2$ norm.

Viewed as a continuum of norms, certain NP-complete problems could be solved by relaxing them to a neighbouring convex equivalent. Tao's proof showed that for these cases the solution to an optimisation of an $L_0$ norm was equivalent to the solution obtained by the $L_1$ norm, and noted that $L_1$ lies on the boundary of convexity. While $L_1$ norms are certainly not linear, their convexity offers the prospect of a polynomial-time algorithm.

\subsection{Soft Thresholding}

Determining the existence of an exactly one-dimensional null-space in a matrix falls into the NP-complete category, since it is related to the existence or absence of a contribution from each singular vector. The problem state is captured by a vector of singular values in which we seek to minimise the number of non-zero entries, so Tao's approach to optimisation can be brought to bear. The algorithmic work following Tao's discovery, chiefly by \cite{osher} and \cite{goldfarb}, centres around \emph{Soft Thresholding}, a simple but non-linear mapping that pushes elements of the solution vector towards zero. A two-step iterative scheme, whose original use appears to have dated back to radar de-noising applications early in the cold war (\cite{bregman}), has been recently re-developed to solve the constrained optimisation, 
\begin{equation}\label{eqn:constrained}
\argmin{\vec{x}} \left\{\|\vec{x}\|_0 \; : \; \vec{y}=\mat{A}\vec{x} \right\},
\end{equation}
governed by some generic linear system by relaxing it to the neighbouring unconstrained convex optimisation, 
\begin{equation}\label{eqn:unconstrainedconvex}
\argmin{\vec{x}} \left\{ \mu\|\vec{y}-\mat{A}\vec{x}\|_2^2 + \|\vec{x}\|_1 \right\}.
\end{equation}
Even the modified problem is not without its challenges, but we present the approach in \S\ref{sec:sparse} using Soft Thresholding.

\subsection{Random projection}

Reorganising the rank-deficiency problem $\mat{B}\vec{\eta}=\vec{0}$ formed from Lie's polynomial series expansion into a suitable linear system is far from straightforward. An approach due to \cite{goldfarb}, organises $\mat{B}$ as a vector (columns stacked on columns) $\mat{B}\to\vec{\hat{b}}$, say, and computes a sequence of scalar products with (pseudo-)random vectors. Using relatively few ($p$) projections of this sort encodes the structure of a vector to a very high probability of uniqueness, and using random vector directions ensures that the projections will not impose any structure of their own on the encoding. The process functions rather like a password-hashing algorithm. 

We then seek to iteratively modify $\vec{\hat{b}}$ so that the corresponding matrix $\mat{B}$ has lower rank, while preserving as much of its original structure as we can. We measure closeness of adherence of our modified matrix $\vec{\hat{b}}_k$ at the $k$'th iteration to the original structure by measuring the discrepancy against the $p\times 1$ vector of scalar products, $\vec{r}=\mat{R}\vec{\hat{b}}$, where $\mat{R}$ is a matrix formed of pseudo-random row vectors. 

The work of \cite{fazelthesis2002} confirms that the convex relaxation of matrix rank is the \emph{nuclear norm} (usually denoted $\|\mat{B}\|_*$), the sum of the singular values. This is simply the $L_1$ norm of the vector of singular values and thus Soft Thresholding is the natural tool to use to find this optimum. The neighbouring convex problem for an $m\times n$ matrix $\mat{B}$ is,
\begin{equation}
\dims{\argmin{\vec{\hat{b}}}}{} \dims{\{ \mu \|}{} \dims{\vec{r}}{^{p \times 1}}\dims{-}{}\dims{\mat{R}}{^{p \times mn}} \dims{\vec{\hat{b}}}{^{mn \times 1}}\dims{\|_2^2 + \|}{} \dims{\mat{B}}{^{m \times n}}\dims{\|_*}{} \dims{\} }{},
\end{equation}
and this equation lies at the core of our algorithm.

\subsection{Numerical inversion of Lie's method}\label{sec:implementoutline}

When iteration has converged, we obtain a modified matrix $\mat{B}_{\infty}$ that is the matrix of smallest obtainable rank when balanced against a Gaussian least-squares best fit to the original matrix $\mat{B}$. Thus the matrix has been de-noised with respect to its rank, and in our original problem rank-deficiency defines the number of `$+c$' degrees of freedom we have for an invariant structure. Lie's method - in its simplest form - seeks just one degree of freedom, so we must successively expand the polynomial basis, de-noise the corresponding linear system, and verify at each step whether $\mat{B}$ has enough columns to have rank-deficiency by one, for a given variance tolerance of the model with respect to data samples as specified by the Lagrange multiplier $\mu$. Once we meet our noise-weighted rank-deficiency criterion, we discard the modified matrix $\mat{B}_{\infty}$, and accept the solution to $\mat{B}\vec{\eta}=\vec{0}$ determines the coefficients of each polynomial basis function, and thus determines the shape of the coordinate transformation required to preserve the invariant structure. The solution to this homogeneous problem is given by $\mat{B}$'s singular value decomposition $\mat{B}=\mat{U}\mat{\Sigma}\mat{V}^T$ and it follows that the elements of $\vec{\eta}$ are given by $\vec{\eta}_i=\mat{V}_{in}$, where $\vec{\eta}$ is an $n\times 1$ vector of unknown coefficients, and the singular values are by convention listed in decreasing order so that the $n$'th column of $\mat{V}$ is the singular vector associated with the least significant singular value. 

Since every row of the matrix $\mat{B}$ represents evaluation of the polynomial basis at one location, then $\mat{B}$ contains a high degree of redundancy. Provided locations are well-distributed across the range of an input data-set, then solutions to the linear system will represent the best-fit projection of the sampled data onto the polynomial basis. The typical differential equation is composed of a small number of polynomial terms, and if the input data-set were noise-free we could expect an exact match and have immediate bijection of the differential equation with its solution. When treating sampled data from real-world measurements as a substitute for a complete analytical solution, we seek to meet the less severe test of best-fit matching of a model over the sampling domain.

\section{A framework for differential equations}\label{sec:liegroups}

The key idea in Sophus Lie's framework for solving differential equations is the notion of symmetry: the ability to change one property of an object, but leave others invariant. We now discuss the crucial geometric insight that allows us to represent integration of differential equations just such a symmetry. Differential equations that take the form,
\begin{equation}
\frac{dx}{dt} = f(t),
\end{equation}
have a solution in terms of an indefinite integral,
\begin{equation}
x= \int f(t) dt + c,
\end{equation}
but Lie's great insight was to reinterpret the constant of integration $c$ as a parameter which slides one solution curve towards another in arbitrarily small increments. If, as indicated in figure \ref{fig:transformationgroup}, the entire $x-t$ plane is filled with valid solution curves that satisfy the differential equation, the parameter $c$ enables us to transform one particular curve into another. If each curve is a member of the \emph{group} of solutions, then the identity element is given by the solution at $c=0$, the inverse is given by $-c$, and provided the plane is infinite in extent, then all transformations remain within the group, so it is closed. Addition of parameters is associative, because $(c+d)+e=c+(d+e)$. This is what we define to be a \emph{Lie group} of solutions to the original differential equation. 

Why is this so profound? If one can find a coordinate transformation that re-expresses an arbitrary differential equation in the form $x(t)=\int f(t) dt + c$, then we can solve the equation in this new coordinate system directly by integration in one variable only. This transfers the problem to finding such a coordinate scheme. The Lie symmetries of the original equation represent directions of transformation that result in no apparent change of the solution (analogous to applying a rotation transformation to a circle). Accordingly, we seek to align an axis of our new coordinate system such that a change in the value of \emph{one} ordinate has no apparent effect on the solution. This ordinate is now oriented in a direction of \emph{invariance} of the solution to the original equation, and so it drops out of any coupling between variables and we are left with integration in a single variable to complete the solution.

What does invariance really mean? It means that we can slide a coordinate variable $x$ to some new position $\hat{x}$, $t$ to $\hat{t}$ and yet still we preserve the relationship between them as given by the original differential equation,
\begin{equation}
\frac{d \hat{x}}{d \hat{t}} = f(\hat{t}\,). 
\end{equation}
This particular differential equation can be integrated directly because the coordinate system already decouples the dependent and independent variables in which the system evolves. This is a special case and we define such an $x-t$ coordinate system to be \emph{canonical coordinates} for this equation. In general, most systems are expressed in terms of coupled variables, and Lie's method is a technique for disentangling these couplings by careful choice of an alternative coordinate system. The simplest non-trivial example would be where $f$ is a function of both $t$ and $x$,
\begin{equation}\label{eqn:nontrivial}
\frac{d x}{d t} = f(t,x),
\end{equation}
and much of the geometric intuition underlying Lie's methods can be developed by representing the equation as a surface  embedded in a space of three ordinates, $t$, $x$ and $\dot{x}$. This space of all coordinates and their relevant derivatives is given the name the \emph{Jet space}. In this geometric context, preservation of the equation's structure as $t\to \hat{t}$, $x\to\hat{x}$ and $\dot{x}\to\hat{\dot{x}}$ is a mapping of this surface to itself. 

\subsection{Jet space}

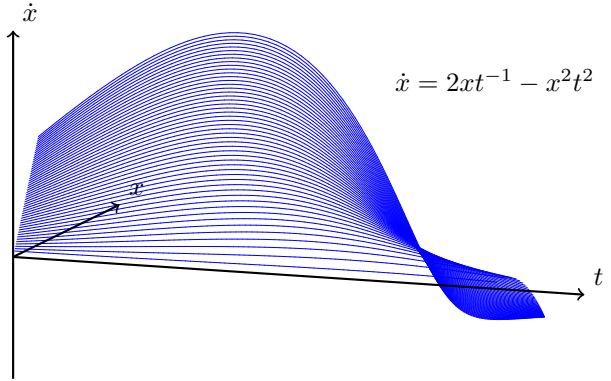
\begin{figure}
\centering
\begin{tikzpicture}

\node[anchor=south west,inner sep=0] (im) at (0,0) {\includegraphics[width=0.9\linewidth]{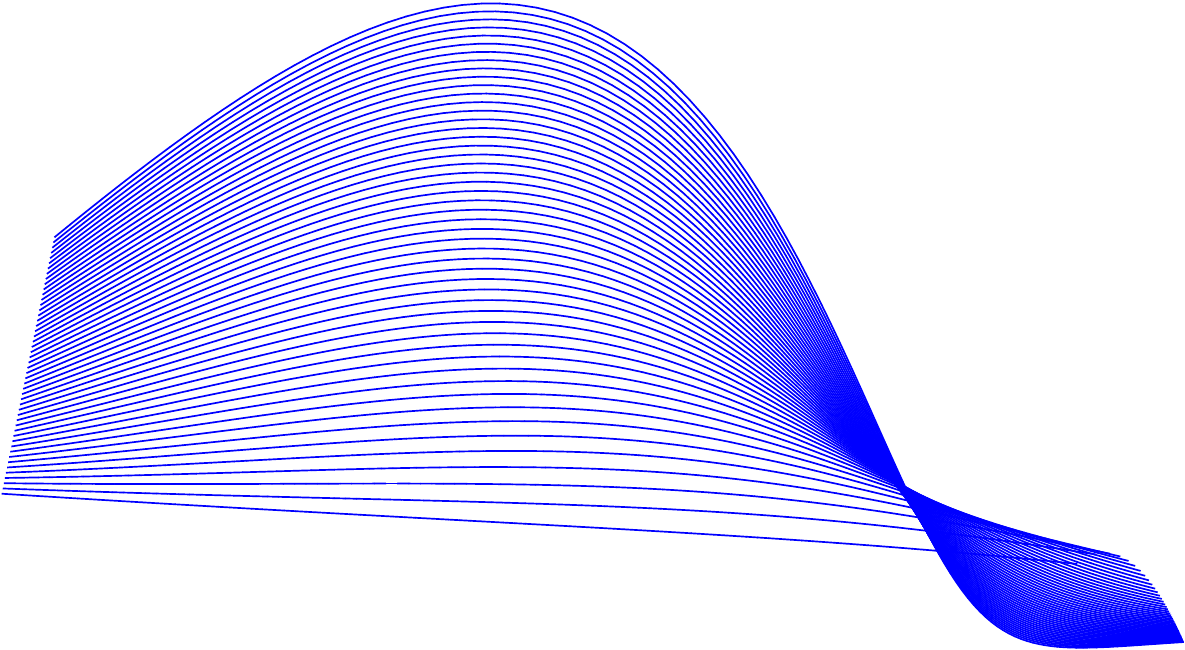}};
\begin{scope}[x={(im.south east)},y={(im.north west)}]

\node[anchor=south] (tlabel) at (1.1,0.09) {$t$};
\node[anchor=south west] (xlabel) at (0.2,0.4) {$x$};
\node[anchor=south west] (xdotlabel) at (0.0,1.0) {$\dot{x}$};
\draw[->,thick] (0.0,0.22) -- (tlabel.south west);
\draw[->,thick] (0.0,0.22) -- (xlabel.south west);
\draw[->,thick] (0.0,-0.2) -- (xdotlabel.south west);

\node[anchor=north west] (eqnlabel) at (0.7,0.9) {$\dot{x}=2xt^{-1}-x^2 t^2$};

\end{scope}
\end{tikzpicture}

\caption{The \emph{Jet space} plot all the variables in a differential equation, including their derivatives to the order found in the equation. In this space, solutions of the equation are trajectories (shown in blue) lying on a manifold of fewer dimensions, given in this instance by $F(t,x,\dot{x})=0=\dot{x}-2\frac{x}{t}+x^2 t^2$.
}
\label{fig:jetspace}
\end{figure}

Geometrically there is no special significance given to the roles of $t$ as an independent variable, $x$ as a dependent variable or $\dot{x}$ as a differential operator, they are simply geometric ordinates for an undulating carpet of points. If we define a function $F(t,x,\dot{x})$ taking arbitrary arguments $t,x,\dot{x}$ as input and returning a value $F$, then locations of constant $F$ are surfaces through the space of $t,x,\dot{x}$. Suppose we choose the function $F$ carefully such that it reflects the behaviour of the differential equation, then one particular iso-surface will define combinations of $t$, $x$ and $\dot{x}$ that satisfy the differential equation. In our particular case the surface given by $F(t,x,\dot{x})=0$ would satisfy the differential equation if,
\begin{equation}
F(t,x,\dot{x})=\dot{x}-f(t,x)=0,
\end{equation}
as shown in figure \ref{fig:jetspace}. Geometric relationships between ordinates $t$, $x$ and $\dot{x}$ are additionally constrained because the rate of change of $x$ with respect to $t$ must equal the value of the $\dot{x}$ ordinate, ie.,
\begin{equation}\label{eqn:xdotconstraint}
\dot{x}=\frac{dx}{dt},
\end{equation}
and this closes the system in the sense that the $x$-values of the surface are both directly dependent on $\dot{x}$ and $t$ (by integrating (\ref{eqn:xdotconstraint}) in $t$) and implicitly defined by $f(t,x)$ by solving $(\ref{eqn:nontrivial})$.

Lie's method for solving differential equations can be considered in two parts: the first part seeks a one-parameter transformation that maps the surface $F=0$ to itself, and the second part then uses this to find an alternative coordinate system $\{u,v,w\}$ aligned with the direction of this one-parameter transformation. Alignnment of one ordinate ensures that the shape of the surface is fully captured by the two remaining ordinates, thus simplifying the problem of integration from three dimensions to two. Often this simpler, decoupled, system may then be integrated directly. For the purposes of equation discovery, our novel approach already has sufficient information at this stage to write down the equation, and the second part of Lie's method is largely redundant.

\subsection{Infinitesimal Generator}

To map the surface $F=0$ to itself, we seek to smoothly slide every point on the surface from their original locations to new ones. We express the original location of an arbitrary point on the surface by a position vector,
\begin{equation}
\vec{p}=\left[ \begin{array}{c} t \\ x \\ \dot{x} \end{array} \right],
\end{equation}
and define a coordinate-sliding parameter $\epsilon$. We then let the position vector $\vec{\hat{p}}(\epsilon)$ represent a new location a distance $\epsilon$ away, and constrain it to remain on $F=0 \;\; \forall \; \epsilon$ and to lie within the vicinity of the identity element of these $\epsilon$-transformations, given by $\vec{\hat{p}}(0)=\vec{p}$.

There is a well-established framework (eg. \cite{moderncontrol}) known as \emph{state-space representation} for describing the evolution of systems in several dimensions. The states of the system are represented as vectors, and the evolution rule for their rate of change may be represented as a linear operator (which can be considered to obey the rules of matrices acting on vectors) that is considered intrinsic to the system. Considering the evolution of $\vec{\hat{p}}$ as a function of $\epsilon$, we may write,
\begin{equation}\label{eqn:statespace}
\frac{d \hat{\vec{p}}}{d \epsilon}=\mat{D}\hat{\vec{p}},
\end{equation}
where $\mat{D}$ is the system's linear operator. Integrating with respect to $\epsilon$ follows the form of the equivalent scalar equation,
\begin{equation}
\hat{\vec{p}}(\epsilon)=e^{\epsilon \mat{D}} \hat{\vec{p}}(0).
\end{equation}
Note that the exponential may be expanded as a Taylor's series,
\begin{equation}\label{eqn:exponentialmap}
e^{\epsilon \mat{D}}=\sum^{\infty}_{k=0} \frac{\epsilon^k \mat{D}^k}{k!} = \mat{I}+\epsilon\mat{D}+\frac{1}{2}\epsilon^2\mat{D}^2+\frac{1}{6}\epsilon^3\mat{D}^3+\hdots \;\;.
\end{equation}
Provided the motion in $\epsilon$ is smooth, then from the geometry we can expand $\vec{\hat{p}}$ as a Taylor's series,
\begin{equation}\label{eqn:vectortaylors}
\vec{\hat{p}}=\vec{p}+\epsilon \frac{d\vec{p}}{d\epsilon}+\frac{1}{2}\epsilon^2\frac{d^2\vec{p}}{d\epsilon^2}+\hdots.
\end{equation}
We now have two separate descriptions or $\vec{\hat{p}}$, (\ref{eqn:statespace}) in terms of some matrix-like linear operator $\mat{D}$, and (\ref{eqn:vectortaylors}) in terms of the differential operator $\frac{d}{d\epsilon}$. By matching terms between their Taylor's expansions, we see immediately that $\mat{D} \equiv \frac{d}{d\epsilon}$. This operator is known as the \emph{infinitesimal generator}. 

\subsection{Tangent space}

\begin{figure}
\centering
\begin{tikzpicture}

\node[anchor=south west,inner sep=0] (im) at (0,0) {\includegraphics[width=0.9\linewidth]{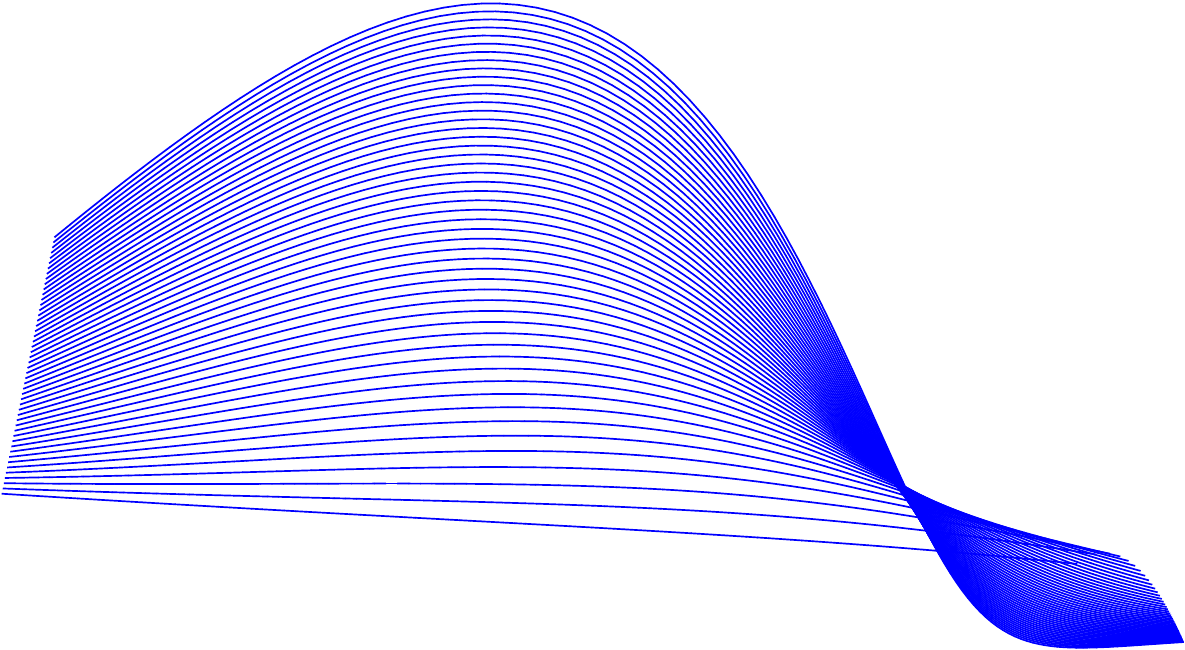}};
\begin{scope}[x={(im.south east)},y={(im.north west)}]

\node[anchor=south] (tlabel) at (1.1,0.09) {$t$};
\node[anchor=south west] (xlabel) at (0.2,0.4) {$x$};
\node[anchor=south west] (xdotlabel) at (0.0,1.0) {$\dot{x}$};
\draw[->,thick] (0.0,0.22) -- (tlabel.south west);
\draw[->,thick] (0.0,0.22) -- (xlabel.south west);
\draw[->,thick] (0.0,-0.2) -- (xdotlabel.south west);

\node[anchor=north west] (delflabel) at (0.73,1.09) {$\del F$};
\node[anchor=south west] (tanglabel) at (0.57,0.65) {};
\node[anchor=west] (plabel) at (0.54,0.72) {$\vec{p}$};
\node[anchor=west] (psimlabel) at (0.72,0.62) {$\overset{\sim}{\vec{p}}$};
\node[anchor=west] (epslabel) at (0.61,0.595) {$\vec{s}$};
\draw[->,ultra thick] (tanglabel.center) -- (delflabel.south west);
\draw[->,ultra thick] (tanglabel.center) -- (0.72,0.58);
\draw[thin,rotate=-42] {(tanglabel.center) ellipse (0.22 and 0.11)};
\draw[thin] (0.0,0.22) -- (tanglabel.center);
\draw[thin] (0.0,0.22) -- (0.72,0.58);

\definecolor{GreyPoint}{rgb}{0.0,0.0,0.0}
\node[anchor=west] (phatlabel) at (0.695,0.49) {$\vec{\hat{p}}$};
\fill[GreyPoint] {(0.69,0.52) ellipse (0.01 and 0.02)};
\draw[thin] (0.0,0.22) -- (0.69,0.52);
\draw[thick,dotted] (0.72,0.58) -- (0.69,0.52);

\node[anchor=north west] (dotp) at (0.7,0.85) {$\vec{s} \cdot \left.\del F\right|_{F=0} =0$};

\end{scope}
\end{tikzpicture}

\caption{The tangent space of the surface is described by a normal vector $\del F$ and any perpendicular vector $\vec{\epsilon}$ lies in the tangent space and points the direction from location $\vec{p}$ to some displaced location $\overset{\sim}{\vec{p}}$ in the tangent plane. Including higher order terms in the Taylor's expansion, we obtain $\vec{\hat{p}}$, a point on the surface of solutions.
}
\label{fig:tangentspace}
\end{figure}

Geometrically, the infinitesimal generator defines a relationship between points on the surface $F=0$. The position vector $\vec{p}$ is smoothly perturbed to a point $\hat{\vec{p}}$. Exactly determining the location $\vec{\hat{p}}$ requires expanding the Taylor's series in $\epsilon$ to arbitrarily high order. We may write the expansion in terms of $t$, $x$ and $\dot{x}$,
\begin{equation}
\left[ \begin{array}{c} \hat{t} \\ \hat{x} \\ \hat{\dot{x}} \end{array} \right] = \left[ \begin{array}{c} t \\ x \\ \dot{x} \end{array} \right] + \epsilon \left[ \begin{array}{c} \frac{\d t}{\d \epsilon} \\ \frac{\d x}{\d \epsilon} \\ \frac{\d \dot{x}}{\d \epsilon} \end{array} \right]+O(\epsilon^2).
\end{equation}
As illustrated in figure \ref{fig:tangentspace}, the higher order terms in the Taylor's expansion account for the `error' between an extrapolation in the plane that is tangent to the surface at $\vec{p}$ and the projection of this tangent-space point onto the `true' curved surface. Locally around $\vec{p}$ there is a one-to-one mapping from the tangent space to the surface, so if we call the point in the tangent space $\overset{\sim}{\vec{p}}$, then the projection to the surface is given by the remaining terms, $k\ge 2$, in the exponential map, 
\begin{equation}
\hat{\vec{p}}=\overset{\sim}{\vec{p}}+\sum^{\infty}_{k=2} \frac{\epsilon^k \mat{D}^k }{k!} .
\end{equation}
If the surface were locally paraboloidal about $\vec{p}$ then including terms only up to second order would be sufficient to describe the exact location $\hat{\vec{p}}$, but an infinity of terms are required to handle the full generality of surface curvatures over an arbitrarily large range of the parameter $\epsilon$. However in practice we need go no further than the highest power of derivative in the original differential equation, so for our simple example we restrict our analysis to first order. 

\subsection{Determining equations}

Algebraically our desire is to find suitable functions $\vec{\hat{p}}(\epsilon)$ such that $F(\vec{\hat{p}})=0 \;\;\forall\;\epsilon$. Using the Taylor's expansion (\ref{eqn:exponentialmap}) enables us to express $F\left(\vec{\hat{p}}\right)$ in terms of $F(\vec{p})$. Provided $F$ has a linearity property $F(\mat{D}\vec{p})=\mat{D}F(\vec{p})$ locally around the identity element $\epsilon=0$, then we have,
\begin{equation}\label{eqn:infinitesimalgenerator}
F\left(\vec{\hat{p}}(\epsilon)\right)=e^{\epsilon\mat{D}} F(\vec{p})=0 .
\end{equation}
It follows immediately that a solution is found if terms at all orders of $\epsilon$ are each zero:
\begin{equation}
F(\vec{p})=0 \;\;,\;\;\mat{D} F(\vec{p})=0\;\;,\;\; \mat{D}^2 F(\vec{p})=0\;\;\hdots\;\;.
\end{equation}
The point $\vec{p}$ is just one of many making up the surface, so the same must be true for \emph{all} points that constitute the surface. These conditions are called the \emph{determining equations}. For a first order differential equation only the terms up to first order in $\epsilon$ in the sequence need to be considered, so the constrains we employ to ensure that $\vec{\hat{p}}$ remains on the surface may be expressed algebraically as, 
\begin{equation}
F=0\;\;,\;\; \frac{d F}{d\epsilon} = 0 .
\end{equation} 

We recognise that $F$ is a function of $t$, $x$ and $\dot{x}$, and using the chain rule we can expand $\mat{D}$ accordingly into partial derivatives,
\begin{equation}
\mat{D}\equiv\frac{d}{d \epsilon}=\frac{d t}{d \epsilon}\frac{\d }{\d t} + \frac{d x}{d \epsilon}\frac{\d }{\d x} + \frac{d \dot{x}}{d \epsilon} \frac{\d }{\d \dot{x}}.
\end{equation}
By expressing the determining equations in partial derivatives, we can identify their geometric consequences. This linear operator acting on $F$ encodes two vectors as a scalar product,
\begin{equation}\label{eqn:dfdescalarproduct}
0=\frac{dF}{d\epsilon}= \Big[\begin{array}{ccc} \frac{d t}{d \epsilon} & \frac{d x}{d\epsilon} & \frac{d \dot{x}}{d\epsilon}\end{array}\Big] \cdot \left[\begin{array}{c} \frac{\d F}{\d t} \\ \\\frac{\d F}{\d x} \\ \\\frac{\d F}{\d \dot{x}} \end{array}\right],
\end{equation} 
one of these is associated directly with the shape of the equation-satisfying surface, and this is the surface normal $\del F$. The other vector (that henceforth we shall call $\vec{s}$) is not yet fully determined, except that the relation $\vec{s}\cdot\del F=0$ mandates that it must lie perpendicular to $\del F$, so as shown in figure \ref{fig:tangentspace} it must lie in the tangent space associated with $\vec{p}$, but there is no further constraint that could produce a unique vector field $\vec{s}(t,x,\dot{x})$. Lie's method addresses this non-uniqueness by requiring that every vector in the field is specified by a common rule, and that rule produces a mapping of the surface to itself that does not become topologically entangled. Rather than imposing such a rule, the method provides a `menu' of options from which the most suitable should be selected. Lie proposes that the menu be composed of polynomial functions of the form $\eta_{ab} t^a x^b$ with a range of powers $a$ and $b$ on offer, and $\eta$ an unknown coefficient associated with a particular polynomial.

In our novel numerical re-interpretation of Lie's method, we recognise that $\del F$ is readily obtainable from the observed, experimental or simulated data-set at each location of interest, and any polynomial can also be evaluated at that location. Since we seek a rule common to all points on the surface, we can simply evaluate the scalar product $\vec{s}\cdot\del F$ at each of them. The general principle can be illustrated by evaluating the product,
\begin{equation}
\begin{split}
&\frac{dt}{d\epsilon}\frac{\d F}{\d t} = \left(\sum \eta_{ab} t^a x^b\right) \frac{\d F}{\d t} = \\
  &\left( \eta_{00} t^0 x^0 + \eta_{01} t^0 x^1 + \eta_{10} t^1 x^0 + \eta_{11} t^1 x^1 + \hdots \right) \cdot \frac{\d F}{\d t},
\end{split}
\end{equation}  
at some particular position $\vec{p}_*$. We split the known values from the unknown polynomial coefficients as follows,
\begin{equation}
\begin{split}
&\left.\frac{dt}{d\epsilon}\frac{\d F}{\d t}\right|_{t_*,x_*} =\\ &\Big[\begin{array}{ccccc} \frac{\d F}{\d t} & \frac{\d F}{\d t} x_* & \frac{\d F}{\d t} t_* & \frac{\d F}{\d t} t_* x_* & \hdots \end{array} \Big]\cdot \left[\begin{array}{c} \eta_{00} \\ \eta_{01} \\ \eta_{10} \\ \eta_{11} \\ \vdots \end{array}\right],
\end{split}
\end{equation} 
and by repeating this for a sample of available positions $\vec{p}_*$, we collect together rows of a matrix $\mat{B}$ of known values calculated from the observed, experimental or simulated data, and isolate a vector of unknown polynomial coefficients $\vec{\eta}$. Once we include all the terms in the expansion of $\frac{dF}{d\epsilon}=0$, we obtain a homogenous linear system, 
\begin{equation}\label{eqn:linearsystem}
\mat{B}\vec{\eta}=\vec{0}. 
\end{equation}

\subsection{Prolongation}\label{sec:prolongation}

We note that the component $\frac{\d \dot{x}}{\d \epsilon}$ is not an independent function, but is constrained by the original differential equation, $\frac{dx}{dt}=f(t,x)$, so we can subtitute for $\dot{x}$ in our equation of surface, ie. $F(t,x,f(t,x))=0$. Since the surface is only a function of $t$ and $x$ the polynomial expansions need only contain these independent variables and this justifies our earlier choice. 

However, there is a dependence of $\dot{x}$ on $t$ and $x$ arising from the constraint (\ref{eqn:xdotconstraint}), and it follows that $\frac{d \dot{x}}{d \epsilon}$ is not independent of $\frac{d x}{d \epsilon}$ and $\frac{d t}{d \epsilon}$. We now seek to obtain an explicit expression coupling the $\dot{x}$ partial derivative to the others, so that the system is fully specified with a minimum number of coefficients $\eta$. 

Our starting point is the Lie symmetry requiring that the differential equation holds at the $\epsilon$-perturbed point $\hat{\vec{p}}$ as well as at $\vec{p}$, so considering the case for $\hat{\vec{p}}$, we have,
\begin{equation}
\hat{\dot{x}}=\frac{d \hat{x}}{d \hat{t}},
\end{equation}
By manipulating the differentials we can express this as a ratio of rates of change with respect to the independent variable $t$,
\begin{equation}\label{eqn:manipdifferentials}
\frac{d \hat{x}}{d \hat{t}}=\frac{\frac{d \hat{x}}{d t}}{\frac{d \hat{t}}{d t}},
\end{equation}
Taking numerator and denominator separately, we can express $\frac{d \hat{x}}{d t}$ and $\frac{d \hat{t}}{d t}$ to first order in $\epsilon$ as,
\begin{equation}\label{eqn:prolongnumden}
\begin{split}
\frac{d \hat{x}}{d t}&=\frac{d }{d t}\left( x+\epsilon\frac{d x}{d \epsilon}\right)=\frac{d x}{d t}+\epsilon\frac{d }{d t}\left(\frac{d x}{d \epsilon}\right)\\&=\frac{d x}{d t}+\epsilon\left(\frac{\d}{\d t}\left(\frac{d x}{d \epsilon}\right) + \frac{d x}{d t}\frac{\d}{\d x}\left(\frac{d x}{d \epsilon}\right)\right) \\ \\
\frac{d \hat{t}}{d t}&=\frac{d }{d t}\left( t+\epsilon\frac{d t}{d \epsilon}\right)=\frac{d t}{d t}+\epsilon\frac{d }{d t}\left(\frac{d t}{d \epsilon}\right)\\&=1+\epsilon\left(\frac{\d}{\d t}\left(\frac{d t}{d \epsilon}\right) + \frac{d x}{d t}\frac{\d}{\d x}\left(\frac{d t}{d \epsilon}\right)\right)
\end{split}
\end{equation} 
It can be shown by binomial expansion that,
\begin{equation}\label{eqn:binomial}
\begin{split}
\frac{1}{(1+\epsilon\alpha)^\beta} =& 1 - \epsilon\alpha\beta + \frac{\beta(\beta+1)}{2!}\epsilon^2\alpha^2 \\&- \frac{\beta(\beta+1)(\beta+2)}{3!}\epsilon^3\alpha^3 +  \hdots,
\end{split}
\end{equation}
and we exploit this to re-express our denominator, considering $\beta=1$ and expanding only to first order in $\epsilon$,
\begin{equation}
\begin{split}
\frac{d \hat{x}}{d \hat{t}}=&\Bigg( \frac{d x}{d t}+\epsilon\left(\frac{\d}{\d t}\left(\frac{d x}{d \epsilon}\right) + \frac{d x}{d t}\frac{\d}{\d x}\left(\frac{d x}{d \epsilon}\right)\right)\Bigg)\\&\times\Bigg( 1 - \epsilon\left(\frac{\d}{\d t}\left(\frac{d t}{d \epsilon}\right) + \frac{d x}{d t}\frac{\d}{\d x}\left(\frac{d t}{d \epsilon}\right)\right)\Bigg).
\end{split}
\end{equation}
Multiplying out the outer brackets but retaining only terms that are first order in $\epsilon$, we obtain,
\begin{equation}
\begin{split}
\frac{d \hat{x}}{d \hat{t}}=&\frac{d x}{d t}+\epsilon\left(\frac{\d}{\d t}\left(\frac{d x}{d \epsilon}\right) + \frac{d x}{d t}\frac{\d}{\d x}\left(\frac{d x}{d \epsilon}\right)\right)\\ &- \frac{d x}{d t} \Bigg( \epsilon\left(\frac{\d}{\d t}\left(\frac{d t}{d \epsilon}\right) + \frac{d x}{d t}\frac{\d}{\d x}\left(\frac{d t}{d \epsilon}\right)\right)\Bigg) \\
=&\frac{d x}{d t}+\epsilon\Bigg(\frac{\d}{\d t}\left(\frac{d x}{d \epsilon}\right) + \frac{d x}{d t}\frac{\d}{\d x}\left(\frac{d x}{d \epsilon}\right)\\&- \frac{d x}{d t}\frac{\d}{\d t}\left(\frac{d t}{d \epsilon}\right) - \frac{d x}{d t}\frac{d x}{d t}\frac{\d}{\d x}\left(\frac{d t}{d \epsilon}\right)\Bigg)
\end{split}
\end{equation}
We make the following connection: from the algebra above we have an explicit expression for $\hat{\dot{x}}$, but according to (\ref{eqn:vectortaylors}), by direct extrapolation from $\vec{p}$ to $\hat{\vec{p}}$ we have an alternative expression,
\begin{equation}
\hat{\dot{x}}=\dot{x}+\epsilon \frac{d \dot{x}}{d \epsilon} + O(\epsilon^2).
\end{equation}
By equating powers of $\epsilon$, then we obtain an explicit expression for $\frac{d \dot{x}}{d \epsilon}$ in terms of the other partial derivatives $\frac{d t}{d \epsilon}$  and $\frac{d x}{d \epsilon}$:
\begin{equation}\label{eqn:dxde}
\begin{split}
\frac{d \dot{x}}{d \epsilon}=&\frac{\d}{\d t}\left(\frac{d x}{d \epsilon}\right) + \dot{x} \frac{\d}{\d x}\left(\frac{d x}{d \epsilon}\right) \\&- \dot{x}\frac{\d}{\d t}\left(\frac{d t}{d \epsilon}\right)  - \dot{x}^2 \frac{\d}{\d x}\left(\frac{d t}{d \epsilon}\right)
\end{split}
\end{equation}

Now we can substitute the polynomial expansions for $\frac{dt}{d\epsilon}$ and $\frac{dx}{d\epsilon}$, which will be of the form,
\begin{equation}\label{eqn:prolongdxde}
\begin{split}
\frac{d \dot{x}}{d \epsilon}=&\frac{\d}{\d t}\left(\sum\eta_{cd} t^c x^d \right)+\dot{x}\frac{\d}{\d x}\left(\sum\eta_{cd} t^c x^d\right)\\&- \dot{x}\frac{\d}{\d t}\left(\sum \eta_{ab} t^a x^b\right)-\dot{x}^2 \frac{\d}{\d x}\left(\sum\eta_{ab} t^a x^b\right),
\end{split}
\end{equation}
where we note that the set of coefficients $\eta_{ab}$ and $\eta_{cd}$ are distinct. Polynomials are straightforward to differentiate, and we obtain,
\begin{equation}
\begin{split}
\frac{d \dot{x}}{d \epsilon}=&\sum c \eta_{cd} t^{c-1} x^d + \dot{x}\sum d \eta_{cd} t^c x^{d-1}\\&-\dot{x}\sum a \eta_{ab} t^{a-1} x^b - \dot{x}^2 \sum b \eta_{ab} t^a x^{b-1}.
\end{split}
\end{equation}
This `prolongation' introduces a coupling into the linear system, because as a result of differentiation, any particular coefficient $\eta$ will in general appear at more than one power of $t$ and $x$. The linear system is organised so each column in $\mat{B}$ associates with one unknown coefficient. Individual elements can expect to receive contributions from two lower-order polynomial basis functions as well as the basis associated with their own column.

\subsection{Linear system}

Homogenous linear systems only have non-trivial ($\vec{\eta}\ne\vec{0}$) solutions if they possess a non-trivial null-space and this occurs when they are rank-deficient. As outlined in \S\ref{sec:implementoutline},  a one-parameter transformation group in $\epsilon$ corresponds to having just one dimension of non-uniqueness in the solution of this linear system. This represents a direction in which coordinates can move without inducing any change in the value of the scalar products $\vec{s}\cdot\del F$, and thus guaranteeing that the transformation is a mapping of the surface to itself. This is precisely the $+c$ condition that motivated Sophus Lie's original insights.

The solution to this homogeneous problem, $\mat{B}\vec{\eta}=\vec{0}$, is given by $\mat{B}$'s singular value decomposition $\mat{B}=\mat{U}\mat{\Sigma}\mat{V}^T$ and it follows that the elements of $\vec{\eta}$ are given by $\vec{\eta}_i=\mat{V}_{in}$, where $\vec{\eta}$ has $n$ elements, and the singular values are by convention listed in decreasing order so that the $n$'th column of $\mat{V}$ is the least significant singular vector.

The difficulty in choosing $\mat{B}$ to have an exactly one-dimensional null space is a considerable challenge in itself, and \S\ref{sec:sparse} is devoted to the details. We progressively increase the richness of polynomial basis that we test against our observed, experimental or simulated data, and for a given tolerance of deviation of the surface mapping over a sample of sufficient coverage, and we cease to expand the polynomial basis once there are enough columns that $\mat{B}$ no longer has full rank. Since each matrix row represents a separate sampled point in the data-set, then for any reasonably-sized data-set the rank of $\mat{B}$ is constrained only by the number of columns - the choice of basis - and not by the number of samples.

\subsection{Closure}

The second part of Lie's method transforms the problem into a new coordinate system $\{u,v,w\}$. The aim is to select an orientation aligned with the vector field $\vec{s}(t,x,\dot{x})$ so that properties remain invariant in the $\vec{s}$ direction and the structure of the system are condensed in to the remaining two variables. Such a convenient axis system is known as \emph{canonical coordinates}. We define coordinate functions $\{\hat{u}(\epsilon),\hat{v}(\epsilon),\hat{w}(\epsilon)\}$ and we treat $\hat{v}(\epsilon)$ as the variable aligned with the vector field $\vec{s}$,  
so $\frac{d \hat{v}(\epsilon)}{d\epsilon}=1$. The remaining two functions $\hat{u}(\epsilon)$ and $\hat{w}(\epsilon)$ should be invariant with respect to $\epsilon$, so their derivatives are zero. Expansion by the chain rule of the coordinate functions $\{u,v,w\}$ into the original coordinates $\{t,x,\dot{x}\}$ produces a set of partial differential equations,
\begin{equation}\label{eqn:canonical}
\left[\begin{array}{c} \frac{d \hat{u}(\epsilon)}{d \epsilon} \\ \\ \frac{d \hat{v}(\epsilon)}{d \epsilon} \\ \\ \frac{d \hat{w}(\epsilon)}{d \epsilon} \end{array}\right]
=
\left[\begin{array}{c} \frac{\d \hat{u}}{\d t}\frac{\d t}{\d \epsilon} + \frac{\d \hat{u}}{\d x}\frac{\d x}{\d \epsilon} + \frac{\d \hat{u}}{\d \dot{x}}\frac{\d \dot{x}}{\d \epsilon} \\ \\ \frac{\d \hat{v}}{\d t}\frac{\d t}{\d \epsilon} + \frac{\d \hat{v}}{\d x}\frac{\d x}{\d \epsilon} + \frac{\d \hat{v}}{\d \dot{x}}\frac{\d \dot{x}}{\d \epsilon} \\ \\ \frac{\d \hat{w}}{\d t}\frac{\d t}{\d \epsilon} + \frac{\d \hat{w}}{\d x}\frac{\d x}{\d \epsilon} + \frac{\d \hat{w}}{\d \dot{x}}\frac{\d \dot{x}}{\d \epsilon}\end{array}\right]=\left[\begin{array}{c} 0\\ \\1\\ \\0\end{array}\right],
\end{equation}
In most cases, sufficient information is already known about the partial differentials to solve for the individual coordinate functions using the method of characteristics, and if one seeks solution of the original differential equation, this may follow by direct integration in the two remaining variables $u$ and $w$. 

In our novel approach, where we attempt the inverse problem of seeking structure in observed, experimental or simulation data, we have no need to explicitly form a $\{u,v,w\}$ coordinate system. In this simplest non-trivial example we have considered, we seek the unknown function $f(t,x)$ from the original differential equation (\ref{eqn:nontrivial}), and by manipulation of the differentials,
\begin{equation}
\frac{d x}{d t}=\frac{\frac{d x}{d \epsilon}}{\frac{d t}{d \epsilon}}=\frac{\sum \eta_{ab}t^ax^b}{\sum \eta_{cd} t^cx^d} =f(t,x),
\end{equation}
we recognise immediately that our polynomial basis functions populate the numerator and denominator, and $f(t,x)$ may be evaluated directly.

\section{Determining dimensionality} \label{sec:sparse}

The core of our new inverse implementation of Lie's method for finding symmetries in differential equations relies on solving a homogeneous linear algebraic system $\mat{B}\vec{\eta}=\vec{0}$ (\ref{eqn:linearsystem}) for a carefully selected matrix $\mat{B}$ whose columns encode polynomial basis functions. These basis functions will describe invariance-preserving transformations of a surface $F(t,x,\dot{x})=0$ mapping to itself and the vector of unknown coefficients $\vec{\eta}$ determines their amplitude. We require $\mat{B}$ to possess a one-dimensional null-space so that it may represent the sole degree of freedom in a one-parameter transformation group. However, when $\mat{B}$ is constructed from noisy observational, experimental or simulated data, there may well be an underlying structure with inter-dependence between columns that becomes masked by noise associated with small singular values that would have otherwise been zero. 

The following discussion builds a framework that will ultimately enable us to perform matrix denoising and address this key step in the inverse problem we seek to solve. In \S\ref{sec:norms}, we discuss the geometry of non-convex optimisation on generic under-determined linear systems of the form $\vec{y}=\mat{A}\vec{x}$, then in \S\ref{sec:convexrelax} we propose a convex relaxation. Between \S\ref{sec:leastsquares} and \S\ref{sec:bregman} we describe the linear algebra required to obtain a computationally feasible algorithm. Finally in \S\ref{sec:lowrank} we return to our motivating problemi of matrix denoising, and adapt the general framework to our specific needs.

\subsection{Optimality of solutions}\label{sec:norms}

Where there is choice or uncertainity amongst a set of valid solutions to a problem, we require some general notion of efficiency to guide our selection. Solutions with arbitrarily large values may be less useful than one with small values relative to the magnitudes of values in the problem, so it is customary to seek a solution that is minimial in some measure of magnitude. Various convenient properties are exhbitied by a least-squares definition of magnitude since there is a useful background of theory connecting it to Euclidian distances and Gaussian statistical distributions, and it often offers closed-form solutions (see \S\ref{sec:leastsquares}). However this is just one special case of a whole class of definitions known as the $L_p$ norms. $L_p$ norms are defined as 
\begin{equation}
L_p(\vec{x})=\left(\sum\limits_i |\vec{x}_i|^p\right)^{\frac{1}{p}}
\end{equation} 
and $p$ can take any positive real value. For $p=2$, the $L_2$ norm, we define magnitude as the root of the sum of the squares of the elements. The $L_\infty$ norm is another popular norm. By raising all the elements of $\vec{x}$ to infinity, this measure of magnitude is dominated by the largest element $|\vec{x}_i|$. In the other limit, $\Lim{p\to 0}$, then the value of any $|\vec{x}_i|$ is only relevant in defining non-zero `existence': $|\vec{x}_i|^0=1 \;\forall \vec{x}_i\ne0$ so the $L_0$ norm simply counts how many values are non-zero.

It turns out that satisfying the $L_0$ norm can produce very useful solutions when trying to identify structure in a problem: these are solutions that use the minimum number of elements in $\vec{x}$ to best fit a solution. This may not be the shortest Euclidian distance from the origin (the $L_2$ norm) but in the sense of \emph{Occam's Razor} it is the least complex solution. If we were to find the $|\vec{x}_i|$ from an $L_0$ minimal solution to a linear system and sort them into decreasing order of magnitude, then for typical real-world data one would expect to  obtain a curve that truncates at the $k$'th sorted element, $k\ll n$, so of $n$ available directions in the linear basis, only $k$ are actually used.

Geometrically, the solution must sit somewhere in the set of $k$-dimensional sub-spaces that are formed from each of the possible combinations of $k$ basis directions. This can be most easily visualised for $n=3$ and $k=2$: here the full space is 3D $\left\{\vec{e_x,e_y,e_z}\right\}$, say, and the sub-spaces are three 2D planes defined, respectively, by $x=0,y=0$ and $z=0$. This is a complicated, very non-linear space in which to look for solutions except by trial and error, and it can be shown that it is an \emph{NP-complete} combinatorical problem (there is no known polynomial-time algorithm). Until recently this seemed like a dead-end, however in 2006 a sequence of publications, \cite{donoho2006,candesrombergjustintao2006a,candesrombergjustintao2006b}, proved that there exists a \emph{convex relaxation} of $L_0$, and this raises the prospect of a tractable (polynomial-time) algorithm. 

The proof can be understood in loose terms by considering the geometry of the $L_p$ norm and comparing it with the geometry of the null-space. The null-space of any matrix $\mat{A}$ is a $(n-k)$-dimensional sub-space. For ease of visualisation, take the example $n-k=2$ and $n=3$, so the null-space is a 2D plane in some general orientation in 3D space. The most familiar norm is the $L_2$, and this represents a Euclidian distance, or radius, from the origin. If we find the point in the null-space with the minimum $L_2$ norm, then we can think of the process as inflating a sphere until it touches the null-space plane. The intersection is the $L_2$ minimum solution. 

The key geometric intuition is that the intersection of an arbitrarily oriented 2D plane with a sphere will to very high probability lie off rather than on a coordinate axis, so solution vectors tend to have non-zero values in many elements. The shape of the $L_p$ ball varies depending on the value of $p$, so for $p>2$ the ball gets corners that become progressively squarer as $p\to\infty$. Going the other way, for $2>p>1$ the sphere gets progressively flattened facets until at $p=1$ the form resembles a rhomboid. For $p<1$ the shape has concave surfaces with protrusions along the coordinate axes. It is in this regime that $L_0$ lives, and being non-convex, there is no guarantee that any local minimum is a global minimum, and posed thus, the problem remains combinatorically difficult. However, \cite{candesrombergjustintao2006a,candesrombergjustintao2006b} realised that any subspace (including the null-space we seek), is `flat' relative to the $L_p$ ball, so if $p<1$ and the $L_p$ ball has protrusions alinged with coordinate axes, as shown in figure \ref{fig:lpball}, intersection with the subspace will be first reached on a protrusion, thus giving the desired \emph{sparse} solution with fewest non-zero entries. So the limiting case $\Lim{p\to 0} L_p$ can be relaxed to $L_{p\le 1}$ and the solution will remain identical. The straight-edged rhomboid obtained when $p=1$ will still intersect the subspace on the same coordinate axis. This is exceptionally convenient, because $L_1$ is the limiting case of a \emph{convex} $L_p$ ball, and thus there is a guarantee that the minimum found by inflating it ifrom the origin is in fact a \emph{global} minimum. Unlike $p<1$ there is no combinatorical explosion of other possibilities to consider. 

\subsection{Problem decomposition}\label{sec:convexrelax}

Ultimately we seek sparse solutions $\vec{x}$ that satisfy the following condition:
\begin{equation}\label{eqn:l0sparse}
\argmin{\vec{x}} \left\{\|\vec{x}\|_0 \; : \; \vec{y}=\mat{A}\vec{x} \right\}.
\end{equation}
but we can relax the condition on $\vec{x}$ to the $L_1$ norm without affecting the sparsity or the correctness of the minimisation. It turns out to require additional work to solve for points that lie \emph{exactly} in the null-space of $\mat{A}$ (see \S\ref{sec:bregman}) so as an first step we relax this condition too, and formulate an unconstrained optimisation of the form,
\begin{equation}\label{eqn:convexrelax}
\argmin{\vec{x}} \left\{ \|\vec{y}-\mat{A}\vec{x}\|_2^2 + \mu\|\vec{x}\|_1 \right\},
\end{equation}
using $\mu$ as a Lagrange multiplier to regularise. However even this is too challenging to solve in one go. Consider instead two simpler sub-problems: 
\begin{equation}
\label{eqn:leastsquaresmin}
\argmin{x} \|\vec{y}-\mat{A}\vec{x}\|_2^2
\end{equation}
and
\begin{equation}
\label{eqn:softthreshold}
\argmin{x} \|\vec{q}-\vec{x}\|_2^2 + \mu\|\vec{x}\|_1, 
\end{equation}
and the following sections provide solutions to those in a form that can be used to reconstitute the original.

\subsection{Least-squares minimisation}\label{sec:leastsquares}

There is a well-known solution to an under-determined, rank-deficient problem of the form (\ref{eqn:leastsquaresmin}). First we define an objective function,
\begin{equation}
\begin{split}
J(\vec{x})&=\left(\vec{y}-\mat{A}\vec{x}\right)^T\left(\vec{y}-\mat{A}\vec{x}\right)\\
          &=\vec{y^T}\vec{y}-2\vec{x^T}\mat{A^T}\vec{y}-\vec{x^T}\mat{A^T}\mat{A}\vec{x},
\end{split}
\end{equation}
that describes the shape of solutions as we vary $\vec{x}$, and $\argmin{\vec{x}}\|\vec{y}-\mat{A}\vec{x}\|_2^2$ will be satisfied at its minimum. It is clear that $J(\vec{x})$ is quadratic in $\vec{x}$, and the minimum is found at,
\begin{equation}
\begin{split}
\frac{\d J}{\d \vec{x}}=0&=-2\mat{A^T}\vec{y} + 2 \mat{A^T}\mat{A}\vec{x} \\
\mat{A^T}\vec{y}&=\mat{A^T}\mat{A}\vec{x} \\
\left(\mat{A^T}\mat{A}\right)^{-1}\mat{A^T}\vec{y}&=\vec{x},
\end{split}
\end{equation}
where we may view the Moore-Penrose pseudo-inverse, $\left(\mat{A^T}\mat{A}\right)^{-1}\mat{A^T}$, as the matrix that most closely represents the inversion of a rank-deficient, non-invertible matrix $\mat{A}$. However straightforward the above may appear in principle, calculating the inversion of $\mat{A^T}\mat{A}$ involves considerable computation, and so this `direct' approach is infeasible for large matrices $\mat{A}$. It turns out that all this can be avoided, and an algorithm can be found where individual operations require only matrix-vector multiplications.

\subsection{Majoration-minimisation}

We now describe a popular technique called \emph{majoration minimisation} for solving equation (\ref{eqn:leastsquaresmin}), replacing a challenging optimisation with something that is locally easier to handle, and performing an iterative procedure in which the minimisation function changes at every iteration. If we call the objective function $J(\vec{x})$, then the `majoration' means that $J(\vec{x})=\argmin{x} \|\vec{y}-\mat{A}\vec{x}\|_2^2$ is replaced by a function $G(\vec{x})$ with two properties. First it must be guaranteed to be everywhere greater than the original $J(\vec{x})$ and so the new function $G(\vec{x})$ satisfies $G(\vec{x}) - J(\vec{x}) \ge 0$. Secondly, the new function must be coincident with the original at the current iteration's estimate of the vector $\vec{x}$, which at the $k$'th iteration is denoted $\vec{x}_k$, so $G\left(\vec{x_k}\right)=J\left(\vec{x_k}\right)$. The hope is that the new function $G(\vec{x})$ is easier to optimise than the original function $J(\vec{x})$, otherwise we have added complexity to the original problem. 

\begin{figure}
\centering
\begin{tikzpicture}

\begin{axis}[clip=false, xmin=-2.9, xmax=2.9, ymin=-1, ymax=12, xlabel=$x$, ylabel=$J(x)$, width=\linewidth, at={(0.52\textwidth,0)}, axis x line=middle, axis y line=middle, height=0.8\linewidth, anchor={south east}]
\addplot[domain=-2.8:3, black, thick] {0.5*((x^4/4)+(x^3/3)-(x^2/2)-x)};

\addplot[domain=-1:3, blue, thick] {0.5*((x^4/4)+(x^3/3)-(x^2/2)-x + 2.5*(x-2.2)^2)};

\definecolor{GreyPoint}{rgb}{0.0,0.0,1.0}
\node[left] at (axis cs:2.2,2.7) {$x_k$};
\node[GreyPoint] at (axis cs:2.2,2.3) {\textbullet};
\draw[-latex,->,dotted] (axis cs:2.2,2.3) -- (axis cs:2.2,0);

\node[left] at (axis cs:1.95,1.2) {$x_{k+1}$};
\node[GreyPoint] at (axis cs:1.5,0.55) {\textbullet};
\draw[-latex,->,dotted] (axis cs:1.5,0.55) -- (axis cs:1.5,0);

\node[left] at (axis cs:1,-0.8) {$x_*$};
\node[GreyPoint] at (axis cs:1,-0.5) {\textbullet};

\node[left] at (axis cs:-0.4,4.5) {$J(x)=\frac{x^4}{8} + \frac{x^3}{6} - \frac{x^2}{4} - x $};
\node[left] at (axis cs:-0.05,6.5) {$G_k(x)=J(x)+\frac{5}{4}(x-\frac{11}{5})^2 $};
\draw[-latex,->,thin] (axis cs:-3.5,3.6) -- (axis cs:-2.8,2.7);
\draw[-latex,->,thin] (axis cs:-3.5,7.4) -- (axis cs:-1,12);

\end{axis}
\end{tikzpicture}

\caption{Majoration-minimisation simplifies optimisation problems by replacing an original objective function $J(x)$ with a function $G_k(x)$ that never lies below $J(x)$. Suitably chosen, it may be easier to compute the stationary point of $G_k(x)$ and iteratively improve until one converges on an optimum $x_*$ that $G_{\infty}(x_*)$ shares with $J(x_*)$.
}
\label{fig:majoration}
\end{figure}
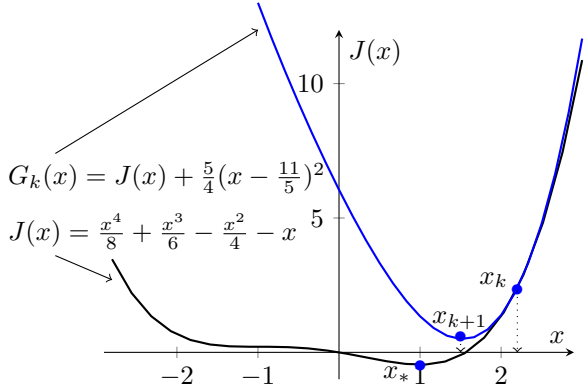

We can always guarantee that the function we choose is everywhere positive by picking a form like $x^2$, and we can guarantee that it lies everywhere above some function $J(x)$ by adding it on, ie. $G(x)=J(x)+x^2>J(x) \;\forall x$. We also must ensure that $G(x_k)=J(x_k)$, and one way to satisfy both conditions is a form $G(x)=J(x)+(x-x_k)^2$. Given that we are considering an iterative scheme and we seeks to converge efficiently, we want the `majoration' to be as small as possible, ie. we want the most slowly growing parabola that still satisfies both conditions. This will ensure that the minimum of $G(x)$ is close to the minimum of $J(x)$ that we ultimately seek, and reduce the number of iterations needed. So $G(x)=J(x)+\alpha(x-x_k)^2$ with a suitably small $\alpha>0$ is an appropriate form of majoration. The key to choosing a computationally efficient $G(x)$ is to carefully pick the majoration so that it cancels out the highest order term(s) in $J(x)$. Figure \ref{fig:majoration} illustrates one iteration of the method.

In the vector case we seek $G(\vec{x})$ whose minimum lies close to but strictly greater than $J(\vec{x}^*)$, than is coincident with $J$ at some $\vec{x_k}$, that is suitably slow-growing around $\vec{x_k}$ to be efficient. For our particular definition of $J(\vec{x})$ the problematic term is $\vec{x^T}\mat{A^T}\mat{A}\vec{x}$, so a well-planned majoration should aim to eliminate this.

\subsection{Landweber iterations}

It turns out that $G_k(\vec{x})$ given by,
\begin{equation}
\argmin{\vec{x}} \left\{ \begin{array}{c} \|\vec{y}-\mat{A}\vec{x}\|_2^2 \\+ \left(\vec{x}-\vec{x}_k\right)\left(a\mat{I}-\mat{A^T}\mat{A}\right)\left(\vec{x}-\vec{x}_k\right) \end{array} \right\}
\end{equation}
satisfies all the requirements of the problem, for suitable $a$ is efficient to converge, and cancels the awkward $\vec{x^T}\mat{A^T}\mat{A}\vec{x}§$ term. The matrix $\left(a\mat{I}-\mat{A^T}\mat{A}\right)$ is positive semi-definite provided $a>\lambda_1\left(\mat{A^T}\mat{A}\right)$, and we need this positivity condition to ensure that the majoration paraboloid has convex curvature in all directions. 
 
\begin{equation}
\begin{split}
G_k(\vec{x})=&\left(\vec{y}-\mat{A}\vec{x}\right)^T\left(\vec{y}-\mat{A}\vec{x}\right)\\&+\left(\vec{x}-\vec{x_k}\right)^T\left(a\mat{I}-\mat{A^T}\mat{A}\right)\left(\vec{x}-\vec{x_k}\right) \\
      =&\vec{y^T}\vec{y} - 2\vec{y^T}\mat{A}\vec{x} + \vec{x^T}\mat{A^T}\mat{A}\vec{x} \\
       &+ \vec{x^T}a\mat{I}\vec{x} - \vec{x^T}a\mat{I}\vec{x_k} \\
       &- \vec{x^T_k}a\mat{I}\vec{x}  + \vec{x_k^T}a\mat{I}\vec{x_k} \\
       &- \vec{x^T}\mat{A^T}\mat{A}\vec{x}   + \vec{x^T}\mat{A^T}\mat{A}\vec{x_k} \\
       &+ \vec{x^T_k}\mat{A^T}\mat{A}\vec{x} - \vec{x^T_k}\mat{A^T}\mat{A}\vec{x_k} 
\end{split}
\end{equation}

We note that the quadratic term, $\vec{x^T}\mat{A^T}\mat{A}\vec{x}$, cancels with this choice of majoration. Noting that $\vec{x_k}$ is a constant throughout each iteration step, remaining terms can be grouped according to their order in $\vec{x}$, 
\begin{equation}
\begin{split}
 G_k(\vec{x})=&\left(\vec{y^T}\vec{y}+a\vec{x^T_k}\vec{x_k}-\vec{x^T_k}\mat{A^T}\mat{A}\vec{x_k}\right)+ \\
        &\left(-2a\vec{x^T}\vec{x_k}+2\vec{x^T}\mat{A^T}\mat{A}\vec{x_k}-2\vec{x^T}\mat{A^T}\vec{y}\right)+ \\
        &\left(a\vec{x^T}\vec{x}\right)
\end{split}
\end{equation}
For the stationary point we obtain,
\begin{equation}
\frac{\d G_k}{\d \vec{x}} = 0 = - 2 a \vec{x_k} + 2 \mat{A^T}\mat{A}\vec{x_k} - 2 \mat{A^T}\vec{y} + 2 a \vec{x},
\end{equation}
and solving for the optimal $\vec{x}$ we arrive at the Landweber iteration:
\begin{equation}\label{eqn:landweber}
\vec{x_{k+1}}=\vec{x_k}+\frac{1}{a}\mat{A^T}\left(\vec{y}-\mat{A}\vec{x_k}\right),
\end{equation}
which requires only two matrix-vector multiplications one by $\mat{A}$ and one by $\mat{A^T}$.

It is worth noting when expanding out the function $G_k(\vec{x})$ that it is quadratic in the optimisation variable $\vec{x}$, because at each iteration $\vec{x_k}$ is considered constant, so $G_k(\vec{x})$ has the form 
\begin{equation}      
\begin{split}
G_k(\vec{x})=p\left(a,\mat{A},\vec{y},\vec{x_k}\right)&-2a\;\vec{q}\left(a,\mat{A},\vec{y},\vec{x_k}\right)^T\vec{x}\\&+ a\vec{x^T}\vec{x}
\end{split}
\end{equation}
with the scalar function $p$ a constant with respect to $\vec{x}$ and the vector function $\vec{q^T}=\vec{x_{k+1}}$ given by the Landweber iteration above, so at the stationary point, 
\begin{equation}
\frac{\d G}{\d x}=0=-2a\vec{q^T}+2a\vec{x^T}
\end{equation}
producing $\vec{x^T}=\vec{q^T}$ as expected. Note that $\|\vec{q}-\vec{x}\|_2^2=\vec{q^T}\vec{q}-2\vec{q^T}\vec{x}+\vec{x^T}\vec{x}$ so excepting the additive constant $\vec{q^T}\vec{q}$ and the multiplicative constant $a$, this behaves like $G_k(\vec{x})$. One implication that can be drawn from this observation is that $G_k(\vec{x})$ has circular level sets.

\subsection{Soft Thresholding} \label{sec:soft}

The second sub-problem (\ref{eqn:softthreshold}), $\argmin{x} \|\vec{q}-\vec{x}\|_2^2 + \mu\|\vec{x}\|_1$, regularises the difference between $\vec{q}$ and $\vec{x}$ with the $L_2$ norm and regularises the vector $\vec{x}$ with the $L_1$ norm - targeting the sparsest solution. Expanding terms, we see that individual elements $\vec{x}_i$ of $\vec{x}$ are de-coupled from each other:  
\begin{equation}
\begin{split} 
J(x)=&(q_1-x_1)^2+\mu|x_1|+ \\
     &(q_2-x_2)^2+\mu|x_2|+ \\
     &(q_3-x_3)^2+\mu|x_3|+...
\end{split}
\end{equation}
 and so we can consider just the scalar case and solve for each element independently.

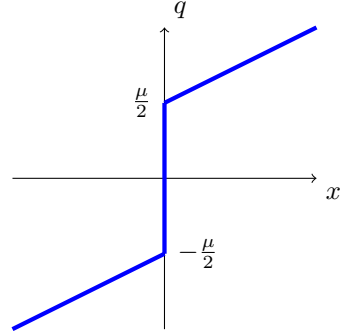
\begin{figure}
\centering
\begin{tikzpicture}

\node[anchor=north west] (xlabel) at (2.0,0) {$x$};
\node[anchor=south west] (ylabel) at (0,2.0) {$q$};
\draw[->] (-2.0,0.0) -- (xlabel.north west);
\draw[->] (0.0,-2.0) -- (ylabel.south west);

\node[anchor=east] at (-0.05,1.0) {$\frac{\mu}{2}$};
\node[anchor=west] at (0.05,-1.0) {$-\frac{\mu}{2}$};
\draw[ultra thick,blue] (-2.0,-2.0) -- (0.0,-1.0);
\draw[ultra thick,blue] (0.0,1.0) -- (2.0,2.0);
\draw[ultra thick,blue] (0.0,-1.0) -- (0.0,1.0);

\end{tikzpicture}

\caption{Soft Thresholding is a non-linear, non-analytic mapping found when minimising functions of the form $f(x)=(q-x)^2 + \mu|x|$.
}
\label{fig:softthresholding}
\end{figure}

If $f(x)=(q-x)^2+\mu|x|$ and we seek $\argmin{x} \left\{f(x)\right\}$ then we seek 
\begin{equation}
\frac{\d f}{\d x} = 0 : f'(x)= 0 = -2(q-x) + \mu\times\sign(x)
\end{equation}
Rearranging for $q$ we have: 
\begin{equation}
q=x+\frac{\mu}{2}\times\sign(x)
\end{equation}
Graphically, this appears as shown in figure \ref{fig:softthresholding}.

and swapping the axes to express $x$ as the dependent variable, we have the following expression for $x$:
\begin{equation}
x=\sign(q) \max\left(0,|q|-\frac{\mu}{2}\right)
\end{equation}
Because elements are de-coupled, this immediately solves the equivalent vector minimisation, and is a non-iterative solution. 

\subsection{Bregman iterations} \label{sec:bregman}

Putting both (\ref{eqn:leastsquaresmin}) and (\ref{eqn:softthreshold}) together, we have, finally,
\begin{equation}
J(\vec{x})=\argmin{\vec{x}} \left\{ \|\vec{y}-\mat{A}\vec{x}\|_2^2 + \mu\|\vec{x}\|_1 \right\}
\end{equation}
 and we solve using majoration-minimisation for a new function $G_k(\vec{x})$. Writing $G_k(\vec{x})$ in the form noted above,
\begin{equation}
G_k(\vec{x})=a\|\vec{q}-\vec{x}\|_2^2 + \mu\|\vec{x}\|_1 + c
\end{equation}
for some constant $c$, and from the Landweber iteration,
\begin{equation}
\vec{q}=\vec{x_k}+\frac{1}{a}\mat{A^T}\left(\vec{y}-\mat{A}\vec{x_k}\right),
\end{equation}
we realise that we can immediately write down the solution to the complete problem by using the solution to (\ref{eqn:softthreshold}) directly,
\begin{equation}
\argmin{x} \left\{ G_k(\vec{x}) \right\} = \sign(\vec{q})\max\left(0,|\vec{q}|-\frac{\mu}{2a}\right)
\end{equation}
 
The Bregman iteration is a rapidly converging procedure to take the unconstrained optimisation
\begin{equation}\label{eqn:unconstrainedL1}
\argmin{\vec{x}} \left\{ \|\vec{y}-\mat{A}\vec{x}\|_2^2 + \mu\|\vec{x}\|_1 \right\}
\end{equation}
and `improve' the enforcement of $\vec{y}=\mat{A}\vec{x}$ such that it becomes a hard constraint rather than a trade-off of regularisation between an $L_1$ norm on $\vec{x}$ and an $L_2$ norm on $\vec{y}-\mat{A}\vec{x}$. The problem statement is then 
\begin{equation}
\argmin{\vec{x}} \left\{ \|\vec{x}\|_1 : \mat{A}\vec{x}=\vec{y} \right\}
\end{equation}

The Bregman algorithm works as a wrap-around for the unconstrained problem (\ref{eqn:unconstrainedL1}) above, and in a form suitable for computation comprises of just two steps:
\begin{equation}
\begin{split}
\vec{x}_{k+1}&=\argmin{\vec{x}} \left\{ \frac{1}{2}\|\vec{y}_k-\mat{A}\vec{x}\|_2^2 + \mu\|\vec{x}\|_1 \right\} \\
\vec{y}_{k+1}&=\vec{y}+(\vec{y}_k-\mat{A}\vec{x}_{k+1})
\end{split}
\end{equation}
This form is known as the 'add-back' form, since it adds the residual error, $\vec{y}_k-\mat{A}\vec{x}_{k+1}$, to the y-vector at each iteration. An unconstrained optimisation must be solved at every Bregman iteration. The Bregman iteration is provably convergent in a finite number of steps (\cite{osher}), and is rapid for suitably chosen parameters $\mu$ and $a$. 

\subsection{Matrix denoising} \label{sec:lowrank}

\begin{figure}
\centering
\begin{tikzpicture}
\node[anchor=south west,inner sep=0] (logos) at (0,0) {
\begin{tabular}{cc}
\includegraphics[width=0.32\linewidth]{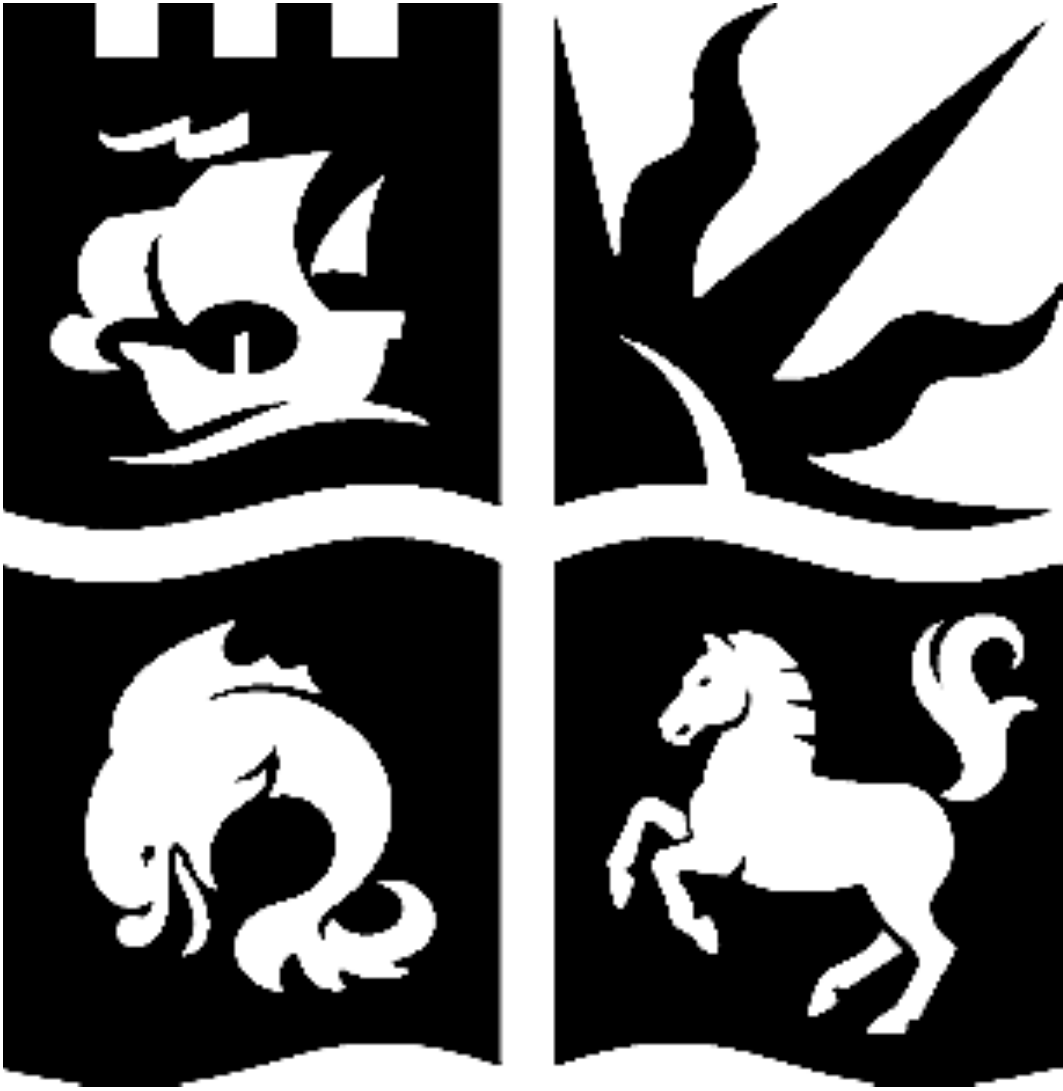} & \includegraphics[width=0.32\linewidth]{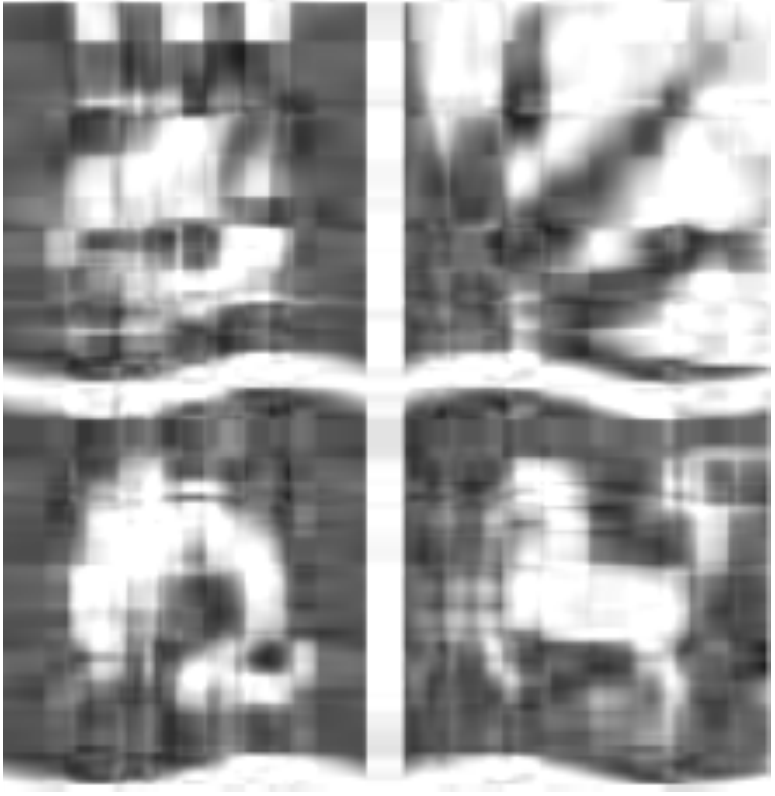} \\
\\
\includegraphics[width=0.32\linewidth]{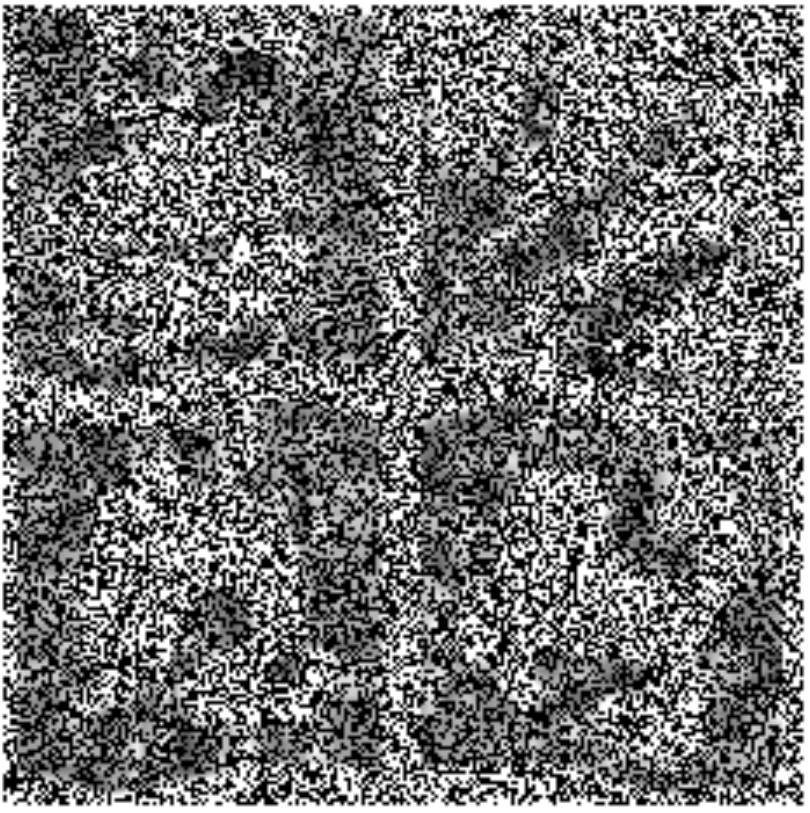} & \includegraphics[width=0.32\linewidth]{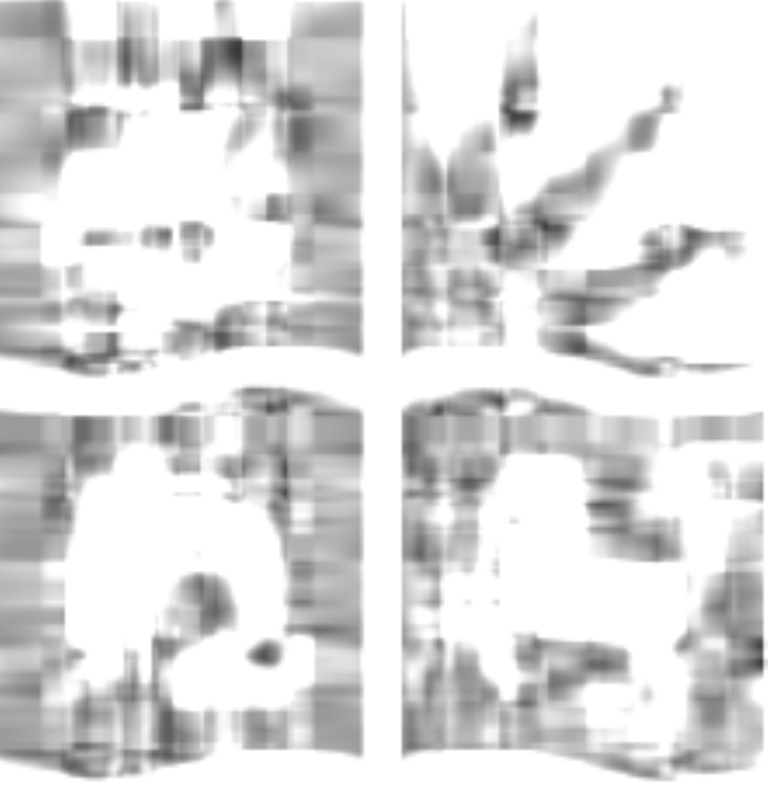}
\end{tabular}
};
\begin{scope}[x={(logos.south east)},y={(logos.north west)}]

\node[anchor=south] (clabel) at (0.25,-0.05) {($c$)};
\node[anchor=south] (alabel) at (0.25,0.48) {($a$)};
\node[anchor=south] (dlabel) at (0.75,-0.05) {($d$)};
\node[anchor=south] (blabel) at (0.75,0.48) {($b$)};

\end{scope}
\end{tikzpicture}
\caption{An example of low-rank structure recovery: the University of Bristol emblem (a), is treated as a matrix $\mat{B}$ (and corresponding image vector $\hat{\vec{b}}$) and rank-reduced to (b), then in (c) random pixels are overwritten, 50\% white, 50\% black. Optimisation problem (\ref{eqn:randoml1l2}) is solved using iterative soft thresholding to recover from (c) a low-rank $\mat{B}$ whose underlying structure is visually recognisable in converged result (d) as closely matching (b).}
\label{fig:rankreduction}
\end{figure}

Reorganising the rank-deficiency problem $\mat{B}\vec{\eta}=\vec{0}$ (\ref{eqn:linearsystem}) formed from Lie's polynomial series expansion into a suitable linear system is far from straightforward. A simple but naive approach might be to manipulate the singular value decomposition of the matrix $\mat{B}=\mat{U}\mat{\Sigma}\mat{V}^T$, so that the vector of singular values $\vec{\sigma}=\mat{\Sigma}\vec{1}$ appears as the unknown in a linear system, ie.,
\begin{equation}
\mat{B}\mat{V}\vec{1} = \mat{U}\vec{\sigma}.
\end{equation}
However, here the singular values retain too little of the structure of $\mat{B}$ to adequately constrain the linear system. A more complete approach, due to \cite{goldfarb}, organises $\mat{B}$ as a vector (columns stacked on columns) $\mat{B}\to\vec{\hat{b}}$, say, and computes a sequence of dot-products with (pseudo-)random vectors. Using relatively few ($p$) projections of this sort encodes the structure of a vector to a very high probability of uniqueness, and using random vector directions ensures that the projections will not impose any structure of their own on the encoding. The process functions rather like a password-hashing algorithm. The random projections form rows of a matrix $\mat{R}$, and the output can be organised as a $p\times 1$ vector $\vec{r}$,
\begin{equation}
\mat{R}\vec{\hat{b}}=\vec{r}.
\end{equation}
We seek to iteratively modify $\vec{\hat{b}}$ so that the corresponding matrix $\mat{B}$ has lower rank, while preserving as much of its original structure as we can. We measure closeness of adherence of our modified matrix $\vec{\hat{b}}_k$ at the $k$'th iteration to the original structure by comparing against the encoded vector $\vec{r}$. We measure,
\begin{equation}
\|\vec{r}-\mat{R}\vec{\hat{b}}_k\|_2^2
\end{equation}
This must be weighted, in the unconstrained convex problem (\ref{eqn:unconstrainedconvex}), with a Lagrange multiplier $\mu$ against some measure of matrix rank. The work of \cite{fazelthesis2002} confirms that the convex relaxation of matrix rank is the \emph{nuclear norm} (usually denoted $\|\mat{B}\|_*$), the sum of the singular values, and this is simply the $L_1$ norm of the vector $\vec{\sigma}$ of singular values and thus Soft Thresholding is the natural tool to use to find this optimum. The neighbouring convex problem for an $m\times n$ matrix $\mat{B}$ is,
\begin{equation}\label{eqn:randoml1l2}
\dims{\argmin{\vec{\hat{b}}}}{} \dims{\{ \mu \|}{} \dims{\vec{r}}{^{p \times 1}}\dims{-}{}\dims{\mat{R}}{^{p \times mn}} \dims{\vec{\hat{b}}}{^{mn \times 1}}\dims{\|_2^2 + \|}{} \dims{\mat{B}}{^{m \times n}}\dims{\|_*}{} \dims{\} }{}.
\end{equation}
Figure \ref{fig:rankreduction} provides a straightforward visual representation of this technique being applied. Treating a 2D image as a matrix, typically its rank is full or nearly full. Truncating an SVD, a low-rank version of the image was produced. By obscuring half of the rank-reduced image with randomly assigned pixel values, we demonstrate recovery of the low-rank image. This illustration is closely related to the Netflix competition (see \cite{netflix} for a review), a competition in computer science to improve the relevance of videos offered to subscribers based on the preferences of previous users with similar tastes. This task can be cast as a matrix completion problem, and with the assumption that relatively few independent parameters govern the choices of large numbers of individuals it can be re-formulated as an $L_0$ minimisation.

\section{Discussion} \label{sec:liedetector}

\subsection{Statistics of $L_p$ norms}

The core of our new algorithm for inverting Lie's method relies on successful denoising of matrix rank, which is an indirect function of all elements of a matrix. One of the reasons convexity is such a highly prized property for constructing robust, efficient algorithms is that optima vary smoothly in the presence of uncertainty, otherwise solutions become combinatorically difficult to find. The statistical properties of these norms thus warrant further discussion. 

We consider the problem posed in (\ref{eqn:l0sparse}), for some unknown $n\times 1$ vector $\vec{x}$. Where the linear system arises from real-world applications, we can expect there to be noise contaminating any underlying signal. Furthermore, we also expect the magnitudes of the elemennts of $\vec{x}$, if they were sorted in decreasing order, to follow a curve. In an noise-free case, we would expect truncation at the $k$'th sorted element, $k\ll n$. In practice noise invariably contaminates a signal, and so there is rarely hard truncation with truly zero elements beyond, but that $n-k$ of them are negligibly small with respect to the dominant few. We thus relax our condition of sparsity to say that the magnitudes of the sorted elements decay according to a power law, ie. $|\vec{x}_i|^{p},\;p<1$. Conveniently, such signals live inside an $L_p$ ball, and so optima discovered should remain unperturbed by noise contamination that satisfies this criterion.

This raises the question of what the unconstrained relaxation (\ref{eqn:convexrelax}) may be doing to the statistics of $\vec{x}$. We consider each element of $\vec{x}$ individually, and model the observed value of the $i$'th element as $x=\chi+w$, where $\chi$ is an underlying signal contaminated by noise $w$. We can pose our question as seeking $\chi$ given $x$, and following \cite{selesnick2008estimation} formulate it as a Bayesian inference problem. Let $\chi_*$ be a \emph{maximum a-posteriori} estimate of $\chi$, such that,
\begin{equation}
\chi_*(x)=\argmax{\chi}\left\{ P(\chi|x) \right\}.
\end{equation}
From the joint probability distribution $P(\chi,x)$, we have the conditional probabilities, $P(\chi|x)=\frac{P(\chi,x)}{P(x)}$ and $P(x|\chi)=\frac{P(\chi,x)}{P(\chi)}$. Bayes' rule gives the relation between them as follows,
\begin{equation}
P(\chi|x)=\frac{P(x|\chi)P(\chi)}{P(x)}.
\end{equation}
Note that the denominator is independent of $\chi$ so $\argmax{\chi}\left\{P(\chi|x)\right\}$ depends exclusively on the numerator, and so we need only seek to maximise the numerator alone.

From statistical thermodynamics $L_2$ norms are known to be closely associated with Gaussian distributions. This relationship arises because entropy (a measure of uncertainty in the system) is maximised by Gaussian distributions when subject to the constraint that the sum of the squares of the element values is an invariant property of that system. This invariance extends to many non-thermodynamic systems, because squared quantities appear in broader contexts, often associated with a fluctuation energy. We consider the distribution of the noise $w$ to be Gaussian with zero-mean, 
\begin{equation}\label{eqn:gaussian}
P(w)=\frac{1}{\sigma_g \sqrt{2 \pi}} e^{\left(-\frac{w^2}{2\sigma_g^2}\right)},
\end{equation}
and from our noise model we note that $P(w)=P(x-\chi)$. It follows that the measurements $x$, given underlying model $\chi$ also follow the same distribution, thus $P(x|\chi)=P(x-\chi)$. Taking this argument and the above recognition that only the denominator is active in the maximisation, then,
\begin{equation}
\chi_*(x)=\argmax{\chi}\left\{ P(x-\chi)P(\chi) \right\}
\end{equation}
To make the problem more tractable we turn this product into an addition by taking logarithms. In general, provided a function $g$ applied to $f(x)$ is monotonic then the transformed output $g(f(x))$ has an extremum at the same location. Formally,
\begin{equation}
\argmax{x} \{f(x)\}=\argmax{x} \{g(f(x))\}\;,\; \d g>0
\end{equation}
Thus we obtain,
\begin{equation}
\chi_*(x)=\argmax{\chi}\left\{ \log\left(P(x-\chi)\right)+\log\left(P(\chi)\right)\right\}
\end{equation}
For a Gaussian distribution of noise (\ref{eqn:gaussian}), we may substitute for $P(x-\chi)$,
\begin{equation}
\chi_*(x)=\argmax{\chi}\left\{ -\frac{(x-\chi)^2}{2\sigma_g^2} + \log\left( P(\chi) \right) \right\},
\end{equation}
leaving the probability distribution of the underlying signal $P(\chi)$ unknown. 

We motivated this discussion of statistical properties by seeking to understand the behaviour of solution to hybrid $L_2$ and $L_1$ minimisation. To extend the analysis we require to make an assumption about the structure of $P(\chi)$. Suppose we choose a Laplacian distribution (whose invariant property is to minimise deviation from the median rather than the mean), 
\begin{equation}
P(\chi)=\frac{1}{\sqrt{2}\sigma_l} e^{-\frac{\sqrt{2}}{\sigma_l} |\chi|},
\end{equation}
which has heavier tails than a Gaussian, then the composed function $\log(P(\chi))$ is given by,
\begin{equation}
\log\left(P(\chi)\right)=-\log\left(\sigma_l\sqrt{2}\right)-\frac{\sqrt{2}}{\sigma_l}|\chi|.
\end{equation}
The stationary point of $\chi_*(x)$ is given by,
\begin{equation}
\frac{\d \chi_*}{\d x} = 0 = \frac{x-\chi}{\sigma_g^2}  -\frac{\sqrt{2}}{\sigma_l}\times \sign(\chi)
\end{equation}
From here is simple to obtain $x$, the observed variable, in terms of the stationary point $\chi_*$,
\begin{equation}
x=\chi_*+\frac{\sqrt{2}\sigma_g^2}{\sigma_l}\times\sign(\chi_*),
\end{equation}
however we seek the inverse relationship, $\chi_*(x)$. Rearranging, we have the maximal \emph{a-posteriori} estimate,
\begin{equation}
\chi_*(x)=\sign(x)\times \textrm{max}\left(0,|x|-\frac{\sqrt{2}\sigma_g^2}{\sigma_l}\right)
\end{equation}
which is precisely the form used to define soft thresholding. Thus we may infer that soft thresholding is equivalent to imposing a Laplacian statistical distribution on the underlying system if the noise is treated as having a Gaussian distribution.  

\subsection{Systems of equations}

There are a number of approaches to generalising from the simplest non-trivial example in (\ref{eqn:nontrivial}) to higher order ODEs, systems of coupled ODEs and to PDEs. The starting point is to recognise that higher order ODEs can always be re-written as an equivalent system of coupled first order equations: for example,
\begin{equation}
t \ddot{x} - 4t^2 \dot{x} - 6x = 0 ,
\end{equation}
can be reformulated by making the substitution $u=\dot{x}$ as,
\begin{equation}\label{eqn:liestatespace}
\left[\begin{array}{c}  \dot{u} \\ \dot{x} \end{array}\right] = \left[\begin{array}{cc} 4t & \frac{6}{t}\\ 1 & 0\end{array}\right]\left[\begin{array}{c}  u \\ x \end{array}\right]. 
\end{equation}
While the form is less condensed than the original differential equation because we introduce an unnecessary degree of freedom (the zero element of the matrix), this system is nonetheless tractable with a small generalisation the first order method laid out in \S\ref{sec:liegroups}. We briefly revisit the algebra, introducing a vector notation for the system of first order ODEs:
\begin{equation}
\frac{d \vec{x}}{dt}=\vec{f}(t,\vec{x}),
\end{equation}
where $\vec{x}$ is the expanded set of state variables and $\vec{f}$ is a set of functions acting on these variables, then without loss of generality we may preserve $t$ as a privileged direction of evolution. As before we seek transformations that permit some new vector $\vec{\hat{x}}$ and some transformed time $\hat{t}$ to be related according to,
\begin{equation}
\frac{d \vec{\hat{x}}}{d \hat{t}}=\vec{f}(\vec{\hat{t},\hat{x}}).
\end{equation}

While difficult to visualise, the Jet space (composed of all coordinate directions and their derivatives) now has an expanded dimensionality. Individually we can write down separate surface equations $F_i(t,\vec{x},\dot{\vec{x}})=0$ for each $i$'th variable. In the above two-variable example,
\begin{equation}
\begin{split}
F_u:\quad 0&=\dot{u}-4tu-\frac{6}{t}x \\
F_x:\quad 0&=\dot{x}-u,
\end{split}
\end{equation}
which both must be simultaneously satisfied. Each equation-satisfying surface $F_i(t,\vec{x},\vec{\dot{x}})=0$ has just one fewer dimensions than the embedding space, so the field of vectors normal to the surface remain uniquely defined and $\epsilon$-sliding transformations are still found by determining equations of the form,
\begin{equation}
F_i(\hat{t},\vec{\hat{x}},\vec{\hat{\dot{x}}})=e^{\epsilon \mat{D}}F_i(t,\vec{x},\vec{\dot{x}})=0,
\end{equation}
where $\mat{D}$ expands by the chain rule across more partial derivatives,
\begin{equation}
\mat{D}=\frac{d}{d\epsilon}=\frac{d t}{d \epsilon} \frac{\d }{\d t} + \sum_i \frac{d \vec{x}_i}{d \epsilon} \frac{\d }{\d \vec{x}_i} + \sum_i \frac{d \vec{\dot{x}}_i}{d \epsilon} \frac{\d }{\d \vec{\dot{x}}_i} .
\end{equation}

The `prolongation' step of \S\ref{sec:prolongation} is more involved with a vector of state variables, since we need to expand $\frac{d \hat{x}_i}{d\hat{t}}$ in terms of un-hatted variables with partial derivatives against each state direction. For each $i$'th direction,
\begin{equation}
\frac{d \vec{\hat{x}}_i}{d \hat{t}}=\frac{\frac{d \vec{\hat{x}}_i}{d t}}{\frac{d \hat{t}}{dt}},
\end{equation}
and the corresponding numerator takes the form,
\begin{equation}
\begin{split}
\frac{d\vec{\hat{x}}_i}{dt} &= \frac{d}{dt}\left( \vec{x}_i+\epsilon\frac{d\vec{x}_i}{d\epsilon}\right) = \frac{d \vec{x}_i}{dt}+\epsilon \frac{d}{dt} \left( \frac{d \vec{x}_i}{d\epsilon}\right) \\
&=\frac{d \vec{x}_i}{dt}+\epsilon \left( \frac{\d}{\d t}\left( \frac{d \vec{x}_i}{d\epsilon}\right) + \sum_j \frac{d\vec{x}_j}{dt} \frac{\d}{\d \vec{x}_j}\left(\frac{d\vec{x}_i}{d\epsilon}\right)\right),
\end{split}
\end{equation}
where partial derivatives are summed over $j$ for each vector component $i$. 
There is a common denominator $\frac{d \hat{t}}{d t}$ that follows the same pattern by replacing the vector component $\vec{x}_i$ with $t$. As observed in (\ref{eqn:prolongnumden}) the first term evaluates as $\frac{dt}{dt}=1$ so using the binomial expansion (\ref{eqn:binomial}), to first order in $\epsilon$, we re-express the denominator in a simple form $1-\epsilon\alpha$. With corresponding functions $\gamma$ and $\delta$ arising from the numerator, we may group terms of common order in $\epsilon$. For the first few orders we obtain,
\begin{equation}
\vec{\hat{\dot{x}}}=\frac{d \vec{\hat{x}}_i}{d \hat{t}}=(\gamma+\epsilon \delta)(1-\epsilon \alpha)=\gamma+\epsilon \delta -\epsilon \alpha \gamma - \epsilon^2 \delta \alpha,
\end{equation}
and comparing with our alternative, geometric Taylor's expansion for $\vec{\hat{\dot{x}}}$, 
\begin{equation}
\vec{\hat{\dot{x}}}=\vec{\dot{x}} + \epsilon \frac{d\vec{\dot{x}}}{d\epsilon}+\hdots\;,
\end{equation}
we obtain the following relations for the low orders in terms of the numerator and denominator functions $\alpha$, $\gamma$ and $\delta$,
\begin{equation}
\begin{split}
\epsilon^0 &: \vec{\hat{\dot{x}}} = \gamma \\
\epsilon^1 &: \frac{d\vec{\dot{x}}}{d\epsilon} = \delta-\alpha \gamma \\
\epsilon^2 &: 0 = -\delta\alpha ,
\end{split}
\end{equation}
so finally we obtain an expression for the coupling between $t, \vec{x}_i$ and $\vec{\dot{x}}_i$ as follows,
\begin{equation}
\begin{split}
\frac{d \vec{\dot{x}}_i}{d\epsilon} =&\frac{\d}{\d t}\left(\frac{\d \vec{x}_i}{\d \epsilon}\right) + \sum_j \frac{d\vec{x}_j}{dt}\frac{\d}{\d \vec{x}_j}\left(\frac{d \vec{x}_i}{d\epsilon} \right)\\& - \frac{d\vec{x}_i}{dt}\left(\frac{\d}{\d t}\left(\frac{d t}{d \epsilon} \right) + \sum_j \frac{d\vec{x}_j}{dt}\frac{\d}{\d \vec{x}_j}\left(\frac{d \vec{x}_i}{d\epsilon} \right)\right)
\end{split}
\end{equation}
Now we substitute a summation over basis functions for $\frac{dt}{d\epsilon}$ and \emph{all} of the $\frac{d\vec{x}_i}{d\epsilon}$, covering each permutation of powers in each of the coordinate directions. Associated with coordinate direction $\vec{x}_j$ and polynomial basis function $i$, we denote the power $b_{ij}$, so every polynomial expansion has the general form,
\begin{equation}
\sum_{i} \left( \eta_i t^{a_i} \prod_j \left( x_j^{b_{ij}}\right) \right),
\end{equation}
with a separate basis function for each permutation of powers $a$ and $b$.

The key step in extending the method is realising that the $+c$ direction on which the coordinate transformation is based must be defined in terms of polynomial basis functions in every coordinate direction in the $\{t,\vec{x},\dot{\vec{x}}\}$ space and must remain consistent with every surface $F_i=0$. This leads to a single, large linear system that encompasses all the constraints, though in principle the complexity of the problem $\mat{B}\vec{\eta}=\vec{0}$ is no greater.

We seek to extend our approach from ordinary to partial differential equations. Typically these may be semi-discretised into a system of similar form to (\ref{eqn:liestatespace}), thus in principle a generalisation of Lie's method to coupled first order systems is sufficient to perform model discovery on higher order systems in multiple dimensions. 

If we were to have prior expectations of the complexity of the system, we could reduce the redundant degree(s) of freedom (the zero(s) in the system matrix) and choose not to offer a polynomial expansion of a variable expected to remain zero. These zeros arise from two sources: firstly as shown in (\ref{eqn:liestatespace}) by the order-reducing substitutions $u=\dot{x}$ and secondly by recognising that physical systems are typically described by local interactions, and these correspond in semi-discrete systems to sparsity in the system matrix, with non-zero values concentrated in a band around the leading diagonal.   

\subsection{Higher order determining equations}

Perhaps a more comprehensive approach is to enforce higher order determining equations,
\begin{equation}
\mat{D}^k F(t,\vec{x},\vec{\dot{x}})=0
\end{equation}
and develop supplementary conditions that must hold in the linear system $\mat{B}\vec{\eta}=\vec{0}$. The operator $\mat{D}^2$ include all the permutations of mixed partial second derivatives, and indeed a full Pascal's triangle of permutations is revealed as further determining equations are taken into account.

The difficulty with this approach is that the Jet space contains all the relevant derivatives of the underlying variables, $\dot{x},\ddot{x},\hdots$, and so the differential operator $\mat{D}$ must be expanded in these variables too. This leads to terms of the form,
\begin{equation}
\frac{d \dot{x}}{d \epsilon}\frac{\d }{\d \dot{x}}\left(\frac{d t}{d \epsilon} \frac{\d F}{\d t} \right).
\end{equation}

Differentiation with respect to $\dot{x}$ of a polynomial expansion is problematic, because the coefficients $\eta$ will themselves determine the rates of change $\frac{dx}{dt}=\dot{x}$, and so there is nonlinearity in this construction of the linear system $\mat{B}\vec{\eta}=\vec{0}$. We note in passing that this difficulty is avoided in any first order case because $\frac{\d }{\d \dot{x}}$ only acts on $F(t,x,\dot{x})$, and this is constant in any given problem. The velocity $\dot{x}$ of observed, experimental or simulated data is readily available by numerical post-processing of the data-set, and the off-solution orientation $\del F$ is computable from the principal curvature of the surface. 

A concise closed form of prolongation at second order may emerge as the least computationally expensive approach to PDE model discovery, but the closure in terms solely of $t$ and $x$ is not obvious. This remains future work.

\subsection{Comparison with neural networks}

Most contemporary research in machine learning focusses on the neural network approach to approximating functions and we introduce the main ideas here before considering the contrast in approach offered by the Lie Detector. The methods developed in the 1980s by Geoffrey Hinton (eg. \cite{hinton1986backprop}) still today form the bedrock of the field. Although inspired by biology, in practice implementations are a small embellishment of ideas familiar from linear algebra. 

Suppose we define connection strengths between neurons to be real-valued parameters, and mandate that every neuron in a layer is connected to every neuron of the next layer and of the previous layer, then every neuron, except those of the input layer, have an activation determined by weighted summation. This can be efficiently encoded as a matrix-vector multiplication: the vector $\vec{a}$ represents the neural activation of the input data, the weights $\mat{W}$ represent the connection strength between the layers, and the vector output $\vec{b}$ represents the activity in the next layer, so $\vec{b}=\mat{W}\vec{a}$. Concatenating lots of layers is expressed easily in the linear algebraic notation; we introduce the notation $\mat{W}^{pq}$ for the weights between some arbitrarily positioned neural layers $\vec{p}$ and $\vec{q}$, and introduce as many as needed. The neural activation through several layers can be calculated as $\vec{z}=\mat{W}^{zy}\mat{W}^{yx} \dots \mat{W}^{bc} \mat{W}^{ab} \vec{a}$. 

The multi-layer model has little utility if each layer is purely a linear algebraic operation, since the single matrix $\mat{\hat{W}}=\prod_{pq} \mat{W}^{pq}$ would supplant the whole sequence. Uniqueness for a particular choice of weight-containing layers is regained by introducing a simple but non-linear saturation \cite{firstuseoftelu2011} that truncates the range of real-valued neuron parameters. Parameters are established by computing the sensitivity (\emph{adjoint}) of a cost function $C$ to each weight $\mat{W}^{pq}_{ij}$ over a set of input data vectors called a \emph{training set}. A local optimum is sought in the parameter space, guided by the gradient vector $-\frac{\d C}{\d W^{pq}_{ij}}$. 

Hinton noticed that using the chain rule, the sensitivity search can be organised into the same structure as the neural network itself, greatly simplifying the task of computing the gradient vector. The vast strides forward (eg. (\cite{lecungradientlearning1998})) since the 1980s have arisen for two interconnected reasons: we can compute much larger networks with modern computer clusters, and algorithmic adjustments have been employed to limit the number of parameters in a network and maintain their sensitivities while still increasing its depth. 

These so-called \emph{convolutional} techniques (\cite{hintonfirstdeeplearning2012}) replace fully-connected networks (matrices dense in non-zero elements) with sparse matrices of an imposed form, flexibility being retained in only $O(9)$ parameters that determine the contribution of certain diagonals. Thus the dimensionality of the optimisation problem grows only linearly with the the depth of the network, and offers the prospect of extensibility limited only by hardware capacity. 

For appropriate choice of diagonals and suitable input data, this approach may be viewed as kernel-filtering, and is related to the action of a linear operator on the data. This has been shown, together with recent developments \cite{generativeadversarial2014,residuallearning2016,kurakin2018adversarial}, to work particularly well in classifying images, where structural information tends to be localised in scale and in position. For example, surface textures tend to be discernible at small scales, image composition is an exclusively large-scale property. 

Despite success with some problems, current methods are not robust to small alterations in images that would pass unnoticed by the human eye. In particular a set of intrinsically similar images that are shifted and rotated to several orientations are not recognised as neighbouring, and brute-force methods that synthetically expand the training set to multiple orientations are currently employed as a workaround. The key difficulty is that there is no intrinsic recognition that the image is invariant with respect to these symmetry transformations. 

Many successful studies performed using neural networks to address image-processing tasks in the medical field, such as \cite{deepretina} on the retina of the eye, and \cite{deepskinmole} on skin moles happen to have intrinsic rotational symmetry, which masks the limitations of a neural network approach. In contrast, our Lie Detector is designed around discovering these very notions of symmetry. 

High dimensional Euclidian spaces (in which multi-megapixel images are simply coordinate locations with position vector $\vec{p}$, say) measure distance between pairs of coordinates $\vec{p}$ and $\vec{\hat{p}}$ as conventional $L_2$ vector norms, $\|\vec{p}-\vec{\hat{p}}\|^2_2$ , and angular orientation remains defined by the scalar product $\vec{p}\cdot\vec{\hat{p}}$. However, both measures of neighbourliness becomes progressively less meaningful as the number of dimensions increases. The values are dependent on the statistical distribution of element values in the position vectors, and for typical data, the central limit theorem indicates how these should converge as the dimensionality increases.  

To very high probability, almost all pairs of vectors will have a small but non-zero scalar product, irrespective of their relative co-alignment. Similarly, Euclidian distances between images become independent of the closeness in overall pattern as human-beings would perceive the same image pair. For example, any image without a regular pattern, shifted in one direction by a pixel, will occupy an entirely different region of the embedding space. The angular alignment given by the pixel-by-pixel correlation will be correspondingly small, and the Euclidian distance largely unrelated to the pattern.
  
High-dimensional spaces are simply too voluminous to sample discretely and expect patterns to emerge by seeking proximity and locality, and neural networks are an attempt to do just that. Our new method seeks to determine continuous symmetries that are valid best-fits across the entire data set, and this will significantly reduce sensitivities on local sampling density. 

The Lie Detector is particularly powerful when applied to simulated data-sets in which the underlying equation set is well-quantified but the emergent behaviour is not, since the simulation may be straightforwardly perturbed in each coordinate variable, computing the local gradients $\frac{\d F}{\d \vec{x}_i}$ to arbitrary accuracy. It is not as straightforward to do so on observed or experimental data of extremely high dimensionality, unless the space is correspondingly well-sampled, or assumptions are made about the expected form of the polynomial basis. 

Analogous to the development of neural networks away from fully connected layers towards sparse convolutional layers, we may without loss of generality restrict the need to seek coefficients for basis functions corresponding to a few diagonals near the leading diagonal of the state matrix. We may also couple the coefficients together as a constant \emph{kernel} (or numerical stencil) over the domain, enforcing zero elsewhere, and constraining the basis functions to act in consort on the system as a linear operator of specified order. 

Boundary regions may be a little more complex and warrant the flexibility of additional basis functions, but coupling and sparsity assumptions would significantly reduce the size of the matrix $\mat{B}$ and the associated computational cost of solving $\mat{B}\vec{\eta}=\vec{0}$ as well as reducing the sampling sensitivities of $\frac{\d F}{\d \vec{x}_i}$. Thus we envisage the Lie Detector having utility on simulated data-sets of arbitrary dimensionality, on inherently well-sampled large-dimensional observational and experimental data with a privileged time-like direction, eg. banking records, land registry data, actuarial risk observations and social media platform archives. 

When sampling densities are inherently lower with respect to the underlying dimensionality, as is the case for observational and experimental configurations, we may recover sufficient population density in the \emph{relevant} region of the Jet space provided the system is in some quasi-steady state. This is particularly important for chaotic systems, and of particular interest to the authors are problems in turbulent fluid flow. 

In a statistically steady state, where states are expected to decorrelate from their initial condition exponentially quickly, suitably spaced instants in time may be considered as independent initial conditions for subsequent evolution trajectories. They then enrich the sampling density over the relevant region until sufficiently well-populated to compute derivatives with confidence. The samples need not be uniformly dense because the best-fit is performed globally across the region of interest.  

\subsection{Improved basis functions}

The polynomial basis in Lie's method acts as a collection of shape functions that may be interpreted as a coordinate transformation that linearise the curvature of the soluton surface. Provided the basis possesses sufficient richness to accommodate the curvature, then the system can be reduced to the form $\mat{B}\vec{\eta}=\vec{0}$ and solved as a linear system. Once the coefficients $\vec{\eta}$ have been determined, the basis functions determine the coordinate transformation associated with a symmetry direction. For the original analytical method, the transformation reduces the dimensionality of the system, simplifiying the problem (often recursively) until it may be solved by direct integration. For the inverse approach we present in this paper, the determination of the coefficients is sufficient to close the problem, since the function $f(t,x)$ (or its matrix equivalent for systems of ODEs) is fully specified by the polynomial expansion. 

The choice of power series polynomials for basis functions is convenient because they are easily differentiable and so the prolongation coupling on $\frac{\d\dot{x}}{\d \epsilon}$ is easily computed. While attractive for solving analytically specified differential equations, some of these features become less important with regard to our inverse methodology.

While positive polynomial indices have been used in the first-principles description of the method, there is no particular case to restrict the basis to positive powers. Introducing negative powers generates a Laurent series and improves the flexibility with which data may be fitted to the basis. Indeed the equation used to create the synthetic data used in our example in this paper contains a negative power of $t$, and a corresponding basis would be a natural fit to the data.

One of the disadvantages of a polynomial basis is that as the power increases (positively or hnegatively) the functions become less distinguishable as linearly independent directions spanning the space of functions, Legendre polynomials can be obtained by a Gram-Schmidt procedure acting on the power series, and over a finite range (spanning the available data-set) have the advantage of being mutually orthogonal, and we would expect a Legendre polynomial basis to have improved robustness. 

Provided the basis is differentiable, any basis can suffice. An obvious alternative choice of basis might be Fourier series instead of polynomials. Orthogonality is guaranteed, function representation is fully general, and differentiation is straightforward. Indeed, there is no particular need to provide the solution in analytical form at all: one simply needs a basis onto which observed, experimental or simulated data may be projected, and these basis functions may be prescribed as look-up tables that themselves may be obtained from observed, experimental or simulated data rather than being encoded as polynomials. Derivatives would be obtained numerically rather than by manipulation of the polynomial power. 
 
\section{Concluding remarks}

The framework Sophus Lie developed for solving differential equations is not especially well known relative to its importance (\cite{howe1983}), and although specific cases are in common usage, the overarching structure is relatively unfamiliar. In fact Lie's ideas have grown to much greater prominence in fields of Physics, where 20th century developments in quantum systems were most naturally expressed using Lie algebras: relationships in the tangent space that are the linear at first order. It has become an important tool for determining orbital states and other symmetries. Einstein's discoveries were made using tensor notation, but it was quickly realised that the subscript-heavy notation was too unwieldy for much further development, and that a description of Riemannian geometry in terms of tangent-spaces and commutators would establish a more convenient framework, and it has become adopted as the modern language for differential geometry. The geometric notions have also found use in robotics, where there is recognition that rotations and translations form a group $SE(n)$ and trajectory optimisation is strongly dependent on curvature of the associated surface. However these contemporary applications exploit the convenient description for curved surfaces and yet none is connected to Lie's original motivations for developing the framework. 

It took until the late 1980s for there to be any revisiting of Lie's work on differential equations. Stephen Wolfram (\cite{mathematica}) developed a computer program, called Wolfram Mathematica, able to manipulate mathematical symbols in the abstract, and codified Lie's methods as outlined in \S\ref{sec:liegroups} for solving differential equations. Wolfram's work was a major step forward in the field and a century after the original thinking, it marks the point at which Lie's framework has come of age and found widespread usage. In Mathematica, the key step in Lie's method, transforming $\mat{D}(F)$ into a linear homogenous equation,
\begin{equation}
\mat{B}\vec{\eta}=\vec{0}
\end{equation}
is performed analytically using automatic algebra, as is the search for a set of basis functions (that form columns of the matrix $\mat{B}$) that is just large enough to admit a one-dimensional null space so that a symmetry direction can be identified. While there are successful approaches, these are cumbersome to encode, and Mathematica truly is a triumph of implementation. 

The Lie Detector we introduce in this paper takes an inverse approach, building the linear system $\mat{B}\vec{\eta}=\vec{0}$ at the core of Lie's framework from observed, experimental or simulated data-sets, de-noising to extract the underlying signal, and discovering the structure of the differential equation. The mathematical insights behind the de-noising have only relatively recently been developed, by Tao and co-workers, and have opened up a field called 'Compressed Sensing'. 

Much initial work in this area has focussed around the so-called `Netflix problem' that can be re-formulated as an $L_0$ minimisation. It turned out, due in part to work by Stanley Osher, that the $L_1$ relaxation has a polynomial-time algorithm. When Osher investigated further, he discovered such technology had been in use - without Tao's rigorous supporting framework - since the 1950s for radar and radio-signal denoising, and the early work of Bregman already provided a reliable and straightforward iterative scheme for finding the $L_1$ norm. 

It is susprising that no-one has previously made the connection between the long-standing availability of de-noising techniques, however weak the underlying justification may originally have been, and a fairly straightforward inversion of Lie's approach for solving differential equations. We surmise that the publicity surrounding Tao's work on relaxation of NP-complete problems brought the problem to the attention of computer scientists, and the incentives offered by the Netflix competition helped accelerate the numerical analysis, but these communities tend not to intersect strongly with those whose main interests lie in analysis of differential equations or geometry. 

We were originally motivated by a desire to extract deeper understanding from data obtained from turbulent flows, where the governing equations are well known but whose evolution is chaotic, and yet on average simple consistent structures emerge. To date the most sophisticated techniques for extracting structure are based on linear eigendecoposition, but this is too restrictive a class of model for most applications and the generality of Lie's framework offers the potential to develop a tool for deriving understanding from arbitrarily general forms of `Big Data'. Brute-force approaches to model discovery using neural networks provide remarkably successful interpolation of existing data, but almost no insight. The insight comes from discovering symmetry: this is the only way to distill structure and it could be argued that this is very definition of understanding.  

In this paper, we have motivated the need for Engineering modelling of complex systems and outlined our approach, for the first time connecting Sophus Lie's long-established framework for differential equations with the modern field of Compressive Sensing that arose from Terence Tao's breakthrough on approximate solutions to NP-complete problems. We then provided a first-principles survey of Lie's method, drawing on the geometric intepretation of these ideas to guide the algebra. Having re-formulated the symmetry-finding problem as a linear system, we detailed the algorithms for removing noise from the system that then facilitate the inversion of Lie's method. We presented our implementation of this inverse procedure and demonstrated successful application to a synthetic data-set in which noise was carefully controlled, and showed that the Lie Detector robustly identifies the underlying symmetries. We then discussed some statistical properties of the technique, extensions to more general classes of model, and conclude with a historical review of the contributing work.

\bibliographystyle{ieeetr}
\bibliography{younglawriebibfile}

\end{document}